\documentclass[letter,12pt,reqno,onside]{amsart}
\usepackage{etex}
\usepackage{graphicx}
\usepackage{setspace}
\usepackage{amsmath}
\usepackage{pdfpages}

\usepackage{amsfonts}
\usepackage{amssymb}
\usepackage{pictex}
\usepackage{amsthm}
\usepackage{enumerate}
\usepackage{graphics,epsfig,verbatim,bm,latexsym,url,amsbsy}
\usepackage{rotating}
\usepackage[authoryear,round]{natbib}
\usepackage{mathrsfs}
\usepackage{array}
\usepackage{url}
\usepackage[top=1in, bottom=1in, left=1in, right=1in, heightrounded,
marginparwidth=2in, marginparsep=2mm]{geometry}

\usepackage{float}
\usepackage{subcaption}
\usepackage{xcolor}
\usepackage{bm}
\usepackage{tikz-cd}
\usepackage{ragged2e}
\usepackage{microtype}
\newcommand{\tr}{\mathsf{T}}
\usepackage{comment}

\theoremstyle{definition} \newtheorem{example}{Example}
\theoremstyle{definition} 
\theoremstyle{definition} \newtheorem{corollary}{Corollary}
\theoremstyle{definition} 
\theoremstyle{definition} 
\theoremstyle{definition} \newtheorem{definition}{Definition}
\theoremstyle{definition} \newtheorem*{PropertyA}{Property A}
\theoremstyle{definition} 
\theoremstyle{definition} \newtheorem{lemma}{Lemma}
\theoremstyle{definition} \newtheorem{theorem}{Theorem}
\theoremstyle{definition} 
\theoremstyle{definition} 
\theoremstyle{definition}\newtheorem{proposition}{Proposition}
\theoremstyle{definition} 
\theoremstyle{definition} 
\theoremstyle{definition} \newtheorem{assumption}{Assumption}
\theoremstyle{definition} 
\theoremstyle{definition} \newtheorem{fact}{Fact}
\theoremstyle{definition} 

\theoremstyle{definition}

\usepackage{multirow}

\long\def\symbolfootnote[#1]#2{\begingroup%
\def\thefootnote{\fnsymbol{footnote}}\footnote[#1]{#2}\endgroup}

\usepackage{footnote}

\usepackage{xr}
\renewcommand{\max}{\operatornamewithlimits{max}}
\renewcommand{\min}{\operatornamewithlimits{min}}

\newcommand{\be}{\begin{equation}}
\newcommand{\ee}{\end{equation}}
\newcommand{\bes}{\begin{equation*}}
\newcommand{\ees}{\end{equation*}}

\newcommand{\rdots}{\mathinner{%
  \mkern1mu\raise1pt\hbox{.}%
  \mkern2mu\raise4pt\hbox{.}%
  \mkern2mu\raise7pt\vbox{\kern7pt\hbox{.}}\mkern1mu}}

\allowdisplaybreaks

\begin{document}

\title[Targeting Interventions in Networks]{Targeting Interventions in Networks}

\author[]{Andrea Galeotti \and Benjamin Golub \and Sanjeev Goyal }


\thanks{We are grateful to the co-editor, Dirk Bergemann, and five anonymous referees for helpful comments. We have also benefited from conversations with Francis Bloch, Drew Fudenberg, Eric Maskin, Matthew Jackson, Asuman Ozdaglar, Francesca Parise, Omer Tamuz, Eduard Talam\`{a}s,  John Urschel, Xavier Vives, Rakesh Vohra, Alex Wolitzky, Leeat Yariv. Ria Granzier-Nakajima, Joerg Kalbfuss, Gustavo Paez, and Eduard Talam\`{a}s provided exceptional research assistance. We thank  Sihua Ding, Joerg Kalbfuss, Fakhteh Saadatniaki, Alan Walsh, and Yves Zenou for detailed comments on earlier drafts.  Andrea Galeotti gratefully acknowledges financial support from  European Research Council through the ERC-consolidator grant (award no. 724356) and the European University Institute through the Internal Research Grant. Benjamin Golub gratefully acknowledges financial support from The Pershing Square Fund for Research on the Foundations of Human Behavior and  the National Science Foundation (SES-1658940, SES-1629446). Galeotti: Department of Economics, London Business School, agaleotti@london.edu. Golub:  Department of Economics, Harvard University, bgolub@fas.harvard.edu; Goyal: Faculty of Economics and Christ's College, University of Cambridge, sg472@cam.ac.uk}

\maketitle

\begin{abstract}
\vspace{-.3in}
We study games in which a network mediates strategic spillovers and externalities among the players. How does a planner optimally target interventions that change individuals' private returns to investment?
We analyze this question by decomposing any intervention into orthogonal \emph{principal components}, which are determined by the network and are ordered according to their associated eigenvalues.  There is a close connection between the nature of spillovers and the representation of various principal components in the optimal intervention. In games of strategic complements (substitutes), interventions place more weight on the top (bottom) principal components, which reflect more global (local) network structure. For large budgets, optimal interventions are simple -- they involve a single principal component.

\end{abstract}

\section{Introduction}

We study games among agents embedded in a network. The action of each agent -- e.g., a level of investment or effort -- directly affects a subset of others, called \emph{neighbors} of that agent.
This happens through two channels: spillover effects on others' incentives, as well as non-strategic externalities. A utilitarian planner with limited resources can intervene to change individuals' incentives for taking the action. Our goal is to understand how the planner can best target such interventions in view of the network and other primitives of the environment.


We now lay out the elements of the model in more detail. Individuals play a simultaneous-move game with continuous actions. An agent's action confers \emph{standalone} benefits on that agent independent of anyone else's action, but it also creates spillovers. The intensity of these spillovers is described by a network, with the strength of a link between two individuals reflecting how strongly the action of one affects the marginal benefits experienced by the other. The effects may take the form of strategic complements or strategic substitutes. In addition to standalone benefits and incentive spillovers, there may be positive or negative externalities imposed by network neighbors on each other.\footnote{This framework encompasses a number of well-known economic examples from the literature: spillovers in educational/criminal effort \citep*{Ballester2006}, research collaboration among firms \citep{GoyalMoraga}, local public goods \citep{BramoulleKranton2007}, investment games and beauty contests \citep{angeletospavan2007,morrisshin2002}, and peer effects in smoking \citep{JacksonRogersZenou2017}.} Before this game is played, the planner can target some individuals and alter their standalone marginal benefits from status quo levels. The cost of the intervention is increasing in the magnitude of the change and is separable across individuals. The planner seeks to  maximize the utilitarian welfare under equilibrium play of the game, subject to a budget constraint.  Our results characterize the optimal intervention policy, showing how it depends on the network, the nature of spillovers, the status quo incentives, and the budget.


An intervention on one individual has direct and indirect effects on the incentives of others. These effects depend on the network and on whether the game features strategic substitutes or complements. For example, suppose the planner increases a given individual's standalone marginal benefits to effort, thereby increasing effort by the targeted individual. If actions are strategic complements, this will push up the incentives of the targeted individual's neighbors. That will increase the efforts of the neighbors of these neighbors, and so forth, creating aligned feedback effects throughout the network. In contrast, under strategic substitutes, the same intervention will discourage the individual's neighbors from exerting effort. However, the effect on those neighbors' neighbors will be positive -- i.e., in the same direction as the effect on the targeted agent. This interplay between spillovers and network structure makes targeting interventions a complex problem. 

At the heart of our approach is a particular way to organize the spillover effects in terms of the \emph{principal components} of the matrix of interactions. Any change in the vector of standalone (marginal) returns can be expressed in a basis of these  principal components. This basis has three special properties: (a) the effects of an intervention along a principal component  is proportional to the intervention, scaled by a network multiplier; (b) the ``network multiplier'' is an eigenvalue of the network corresponding to that principal component; (c) the principal components are orthogonal, so that the effects along various principal components can be treated separately (in a suitable sense). These properties allow us to express the effect of interventions on actions and welfare in a way amenable to a parsimonious characterization of optimal interventions.

Our main result, Theorem 1, characterizes the optimal intervention in terms of how similar it is to various principal components -- or, in other words, how strongly represented various principal components are in it.\footnote{We use the standard notion of cosine similarity: the similarity of two vectors is the cosine of the angle between them in the plane they jointly define.} Building on this characterization, Corollary 1 describes how the representation of various principal components in the optimal intervention is shaped by the  nature of the strategic interaction. The principal components can be ordered by their associated eigenvalues (from high to low). In games of strategic complements, the optimal intervention is most similar to the first principal component -- the familiar \emph{eigenvector centrality} -- and progressively less similar as we move down the principal components. In games of strategic substitutes, the order is reversed: the optimal intervention is most similar to the \emph{last} principal component.  The ``higher'' principal components capture the more global structure of the network: this is important for taking advantage of the aligned feedback effects arising under strategic complementarities. The ``lower'' principal components capture the local structure of the network: they help the planner to target the intervention so that it does not cause crowding out between adjacent neighbors: this is an important concern when actions are strategic substitutes.

We then turn to the study of \textit{simple} optimal interventions, i.e., ones where the intervention (change in the standalone marginal benefit) for each node is determined by a single network statistic of that node, and invariant to other primitives such as status quo incentives. Propositions 1 and 2 show that for large enough budgets the optimal intervention is simple: in games of strategic complements, the optimal intervention vector is proportional to the first principal component, while in games of strategic substitutes, it is proportional to the last one.\footnote{In similarity terms, this means that the optimal intervention has a cosine similarity of nearly $1$ to the first or last principal component (depending on the case), and a similarity of nearly $0$ to all other principal components.} Moreover, the network structure determines how large the budget must be for optimal interventions to be simple. In games of strategic complements (substitutes), the important statistic is the gap between the top (bottom) two eigenvalues. When this gap is large, even at moderate budgets the intervention is simple.

Theorem 1 obtains in a setting where the planner knows the status quo standalone marginal benefits of all individuals. Our methods can also be used to study optimal interventions in a setting where the planner does not know these benefits but knows only their distribution. In such a setting, the impact of an intervention on expected social welfare is determined by how it alters the first and second moments of the benefits distribution. Propositions 3 and 4 characterize optimal interventions and show that the key insights about the order of targeting of principal components extend.

We now place the paper in the context of the literature. The intervention problem we study concerns optimal policy in the presence of externalities. Research over the past two decades has deepened our understanding of the empirical structure of networks and the theory of how networks affect strategic behavior.\footnote{See, for example, \citet*{Goyaletal2006}, \citet*{Ballester2006}, \citet*{bramoulle2014strategic}, and \cite*{Galeottietal2010}} This has led to the study of how policy design should incorporate information about networks. Network interventions are currently an active subject of research not only in economics but also in related disciplines such as computer science, sociology and public health.\footnote{For a general introduction to the subject, see \cite{Rogers1983}, \cite*{Kempeetal2003}, \cite{Borgatti2006}, and \cite{Valente2012}. Within economics, a prominent early contribution is \citet*{Ballester2006}; recent contributions include \citet*{banerjee2013diffusion}, \cite{BelhajDeroian}, \cite{BlochQuerou2013}, \citet*{Candoganetal2012}, \cite{GDemange}, \cite{FainmesserGaleotti2017}, \cite{GaleottiGoyal2009}, \cite{GaleottiRogers}, \cite*{leduc2017pricing}, and \citet*{akbarpour}.} The main contribution of this paper is methodological. It lies in (i) using the principal components approach to decompose the effect of an intervention on social welfare and (ii) using the structure afforded by this decomposition to characterize optimal interventions. Of special interest is the close relation between the strategic structure (complements or substitutes) and the appropriate principal components to target.\footnote{Online Appendix Section \ref{OA-sec:othermeasures} presents a discussion of the relationship between principal components and other network measures that have been studied in the literature.}

The rest of the paper is organized as follows. Section \ref{sec:basic} presents the optimal intervention problem.
Section \ref{sec:PC} sets out how we apply a principal component decomposition to our game. Section \ref{sec:optimal_targetS} characterizes  optimal interventions.  Section \ref{sec:incompleteinformation} studies a setting where the planner has incomplete information about agents' standalone marginal returns. Section \ref{sec:concludingremarks} concludes. Appendix A contains the proofs of the main results. The Online Appendix presents the proofs of Propositions 3 and 4 and discusses a number of extensions.

\section{The model}\label{sec:basic}

We consider a simultaneous-move game among individuals $\mathcal{N}=\{1,\ldots,n\}$, where $n\geq 2$. Individual $i$ chooses an action, $a_i \in \mathbb{R}$. The vector of actions is denoted by $\bm{a} \in \mathbb{R}^n$. The payoff to individual $i$ depends on this vector, $\bm{a}$, the \emph{network} with adjacency matrix $\bm{G}$, and other parameters, described below:
\begin{equation}
\label{payoffs}
U_i(\bm{a},\bm{G})=\underbrace{a_i\left(b_i+\beta\sum_{j}g_{ij}a_j\right)}_{\text{returns from own action}} -\underbrace{\frac{1}{2}a_i^2}_{\substack{\text{private costs} \\ \text{of own action}}}+\underbrace{P_i(\bm{a}_{-i},\bm{G},\bm{b})}_{\text{pure externalities}}.
\end{equation}	

The private marginal returns from increasing the action $a_i$ depend both on $i$'s own action, $a_i$, and on others' actions. The coefficient $b_i \in \mathbb{R}$ corresponds to the part of $i$'s marginal return that is independent of others' actions, and is thus called $i$'s \emph{standalone marginal return}.  The contribution of others' actions to $i$'s marginal return is given by the term $\beta \sum_j g_{ij} a_j$. Here  $g_{ij} \geq 0$ is a measure of the strength of the interaction between $i$ and $j$. The parameter $\beta$  captures strategic interdependencies. If $\beta>0$, then actions are strategic complements; if $\beta<0$, then actions are strategic substitutes. The function $P_i(\bm{a}_{-i},\bm{G},\bm{b})$ captures \emph{pure externalities} -- that is, spillovers that do not affect the best response. The first-order condition for individual $i$'s action to be a best response is:
\begin{equation*}\label{bestresponse}
a_i=b_i+\beta\sum g_{ij}a_j.
\end{equation*}
Any Nash equilibrium action profile $\bm{a}^*$ of the game satisfies
\begin{equation}\label{LS}
[\bm{I}-\beta \bm{G}]\bm{a}^*=\bm{b}.
\end{equation}

We now make two assumptions about the network and the strength of strategic spillovers. Recall that the spectral radius of a matrix is the maximum of its eigenvalues' absolute values.
\begin{assumption}\label{symmetry}
The adjacency matrix $\bm{G}$ is symmetric.\footnote{We extend our analysis to more general $\bm{G}$ in the Online Appendix Section \ref{OA-sec:R1}.}
\end{assumption}

\begin{assumption}\label{as:spectral_radius}
The spectral radius of $\beta \bm{G}$  is less than $1$,\footnote{An equivalent condition is for $|\beta|$ to be less than the reciprocal of the spectral radius of $\bm{G}$.} and all eigenvalues of $\bm{G}$ are distinct (the latter condition holds generically).
\end{assumption}
\noindent Assumption \ref{as:spectral_radius} ensures that (\ref{LS}) is a necessary and sufficient condition for a solution, and also ensures the uniqueness and stability of the Nash equilibrium.\footnote{See \citet{Ballester2006} and \citet{bramoulle2014strategic} for detailed discussions of this assumption and the interpretation of the solution given by (\ref{eq:LS-solved}).} 
Under these assumptions, the unique Nash equilibrium of the game can be characterized by
\begin{equation}
\bm{a}^* = [\bm{I}-\beta \bm{G}]^{-1} \bm{b}. \label{eq:LS-solved}
\end{equation}

The utilitarian social welfare at equilibrium is given by the sum of the equilibrium utilities:
\[
W(\bm{b},\bm{G})=\sum_i U_i(\bm{a}^*,\bm{G}).
\]
The planner aims to maximize the utilitarian social welfare at equilibrium. She does this by changing the status quo standalone marginal returns $\hat{\bm{b}}$, to new values, ${\bm{b}}$, subject to a budget constraint on the cost of her intervention. The timing is as follows. The planner moves first and chooses her intervention, and then individuals simultaneously choose actions. The incentive-targeting (IT) problem is given by
\begin{align*}\label{QG:P1}
\max_{\bm{b}} &  \text{ } W(\bm{b},\bm{G}) \tag{IT}\\
\text{s.t.:} \text{   } & \bm{a}^* = [\bm{I}-\beta \bm{G}]^{-1} \bm{b}, \nonumber\\
 K(\bm{b}, \hat{\bm{b}}) &=  \sum_{i \in \mathcal{N}}\left(b_i-\hat{b}_i\right)^2\leq C, \nonumber
\end{align*}
where $C$ is a given budget. The marginal costs of altering the $b_i$ are separable across individuals and increasing in the magnitude of the change for each individual. For discussions and extensions on more general cost functions, see the Online Appendix Section \ref{OA-sec:R3}.
In the Online Appendix Section \ref{OA-sec:R4}, we study a setting in which a planner provides monetary payments to individuals that induce them to change their actions. We show that the resulting optimal intervention problem has the same mathematical structure as the one we study in our basic model.

We present two economic applications to illustrate the scope of our model. The first example is a classical investment game, and the second example is a game of providing a local public good.

\noindent\begin{example}[The investment game]\label{Ex1}

Individual $i$ makes an investment $a_i$ at a cost $\frac{1}{2}a_i^2$. The private marginal return on that investment is $b_i+\beta\sum_{j}g_{ij}a_j$, where $b_i$ is individual $i$'s standalone marginal return and $\sum_{j}g_{ij}a_j$ is the aggregate local effort. The utility of $i$ is
\[
U_i(\bm{a},\bm{G})=a_i \left( b_i+\beta\sum_{j}g_{ij}a_j \right)-\frac{1}{2}a_i^2.
\]
The case with $\beta>0$ is the canonical case of investment complementarities as in \citet{Ballester2006}. Here, an individual's marginal returns are enhanced when his neighbors work harder; this creates both strategic complementarities and positive externalities. The case of $\beta<0$ corresponds to strategic substitutes and negative externalities; this can be microfounded via a model of competition in a market after the investment decisions $a_i$ have been made, as in \citet{GoyalMoraga}.

It can be verified that the equilibrium utilities, $U_i(\bm{a}^*,\bm{G})$, and the utilitarian social welfare at equilibrium, $W(\bm{b},\bm{G})$, are as follows:
\[
U_i(\bm{a}^*,\bm{G})=\frac{1}{2}(a^{*}_i)^2  \text{ and } W(\bm{b},\bm{G})=\frac{1}{2}\left( \bm{a}^* \right)^{\tr} \bm{a}^*.
\]

\end{example}

\noindent
\begin{example}[Local public good]\label{Ex4}
Following  \citet{BramoulleKranton2007}, \citet{GaleottiGoyal2010}, and \citet{allouch2015private,allouch2017}, we consider a local public goods problem -- for instance, collecting non-rival information. Without information-acquisition costs, the optimal amount of information to acquire would be $\tau$.\footnote{This can be taken to be the maximum amount of information available; equilibrium acquisitions will always be less than this.} Individual $i$ has an amount $\tilde{b}_i<\tau$ of information to begin with. He can expend effort to personally acquire additional information, increasing his amount of information to $\tilde{b}_i+a_i$. If his neighbors acquire information, then he can also access $\tilde{\beta}\sum_j g_{ij}a_j$, with $\tilde{\beta}\in(0,1]$ capturing a loss in the transmission of information. The total information that individual $i$ has is
\[
x_i=\tilde{b}_i+a_i+\tilde{\beta}\sum_j g_{ij}a_j.
\]
The utility of $i$ is
\[
U_i(\bm{a},\bm{G})=-\frac{1}{2}(\tau-x_i)^2-\frac{1}{2}a_i^2.
\]
This is a game of strategic substitutes and positive externalities.  Performing the change of variables $b_i=[\tau-{b}_i]/2$ and $\beta=-\tilde{\beta}/2$ (with the status quo equal to $\hat{b}_i=[\tau-\tilde{b}_i]/2$)  yields a best-response structure exactly as in condition (\ref{LS}). The aggregate equilibrium utility is $W(\bm{b},\bm{G})=- \left( \bm{a}^* \right)^{\tr}\bm{a}^*$.

\end{example}

These two canonical examples share a technically convenient property:

\begin{PropertyA}\label{Ass:payoff}
The aggregate equilibrium utility is proportional to the sum of the squares of the equilibrium actions, that is, $W(\bm{b},\bm{G})=w \cdot \left( \bm{a}^* \right)^{\tr}\bm{a}^*$ for some $w\in\mathbb{R}$, where $\bm{a}^*$ is the Nash equilibrium of the network game.
\end{PropertyA}

Online Appendix Section \ref{OA-sec:PureBC}  discusses a network beauty contest game inspired by \citet{morrisshin2002} and \citet{angeletospavan2007}  which also satisfies this property. While Property A facilitates analysis, it is not essential. Online Appendix Section \ref{OA-sec:R2} extends the analysis to cover important cases where  this property does not hold.



\section{Principal components}\label{sec:PC}

This section introduces a basis for the space of standalone marginal returns and actions in which, under our assumptions on $\bm{G}$, strategic effects and the welfare function of interest to the planner both take a simple form.

\begin{fact} \label{fact:lambda}
If $\bm{G}$ satisfies Assumption \ref{symmetry}, then $\bm{G}=\bm{U}\bm{\Lambda} \bm{U}^\tr$,  where:
\begin{enumerate}[(1)]
\item[1.] $\bm{\Lambda}$ is an $n \times n$ diagonal matrix whose diagonal entries $\Lambda_{ll}=\lambda_l$ are the eigenvalues of $\bm{G}$ (which are real numbers), ordered from greatest to least: $\lambda_1 \geq \lambda_2 \geq \cdots \geq \lambda_n$.
\item[2.] $\bm{U}$ is an orthogonal matrix. The $\ell^{\text{th}}$ column of $\bm{U}$, which we call $\bm{u}^\ell$, is a real eigenvector of $\bm{G}$, namely the eigenvector associated to the eigenvalue $\lambda_{\ell}$, which is normalized in the Euclidean norm: $\Vert \bm{u}^\ell \Vert=1$.
\end{enumerate}
\end{fact}
For generic $\bm{G}$, the decomposition is uniquely determined, except that any column of $\bm{U}$ is determined only up to multiplication by $-1$.

An important interpretation of this diagonalization is as a decomposition into \emph{principal components}. We can think of the columns of $\bm{G}$ as $n$ data points. The first principal component of $\bm{G}$ is defined as the $n$-dimensional vector that minimizes the sum of squares of the distances to the columns of $\bm{G}$. The first principal component can therefore be thought of as a fictitious column that ``best summarizes'' the dataset of all columns of $\bm{G}$. To characterize the next principal component, we orthogonally project all columns of $\bm{G}$ off this vector and repeat this procedure for the new columns. We continue in this way, projecting orthogonally off the (subspace generated by) vectors obtained to date, to find the next principal component. A well-known result is that the eigenvectors of $\bm{G}$ that diagonalize the matrix (i.e., the columns of $\bm{U}$) are indeed the principal components of $\bm{G}$ in this sense. Moreover, the eigenvalue corresponding to a given principal component quantifies the residual variation explained by that vector.  

\begin{figure}
			\includegraphics[scale=0.22]{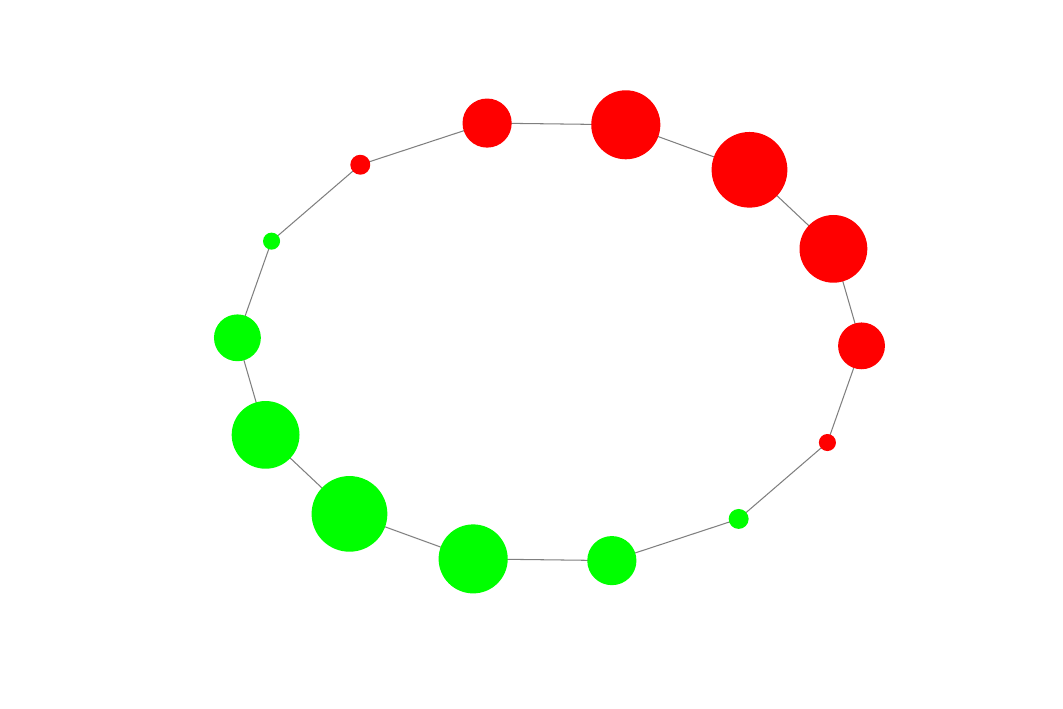}
		\;\;\includegraphics[scale=0.22]{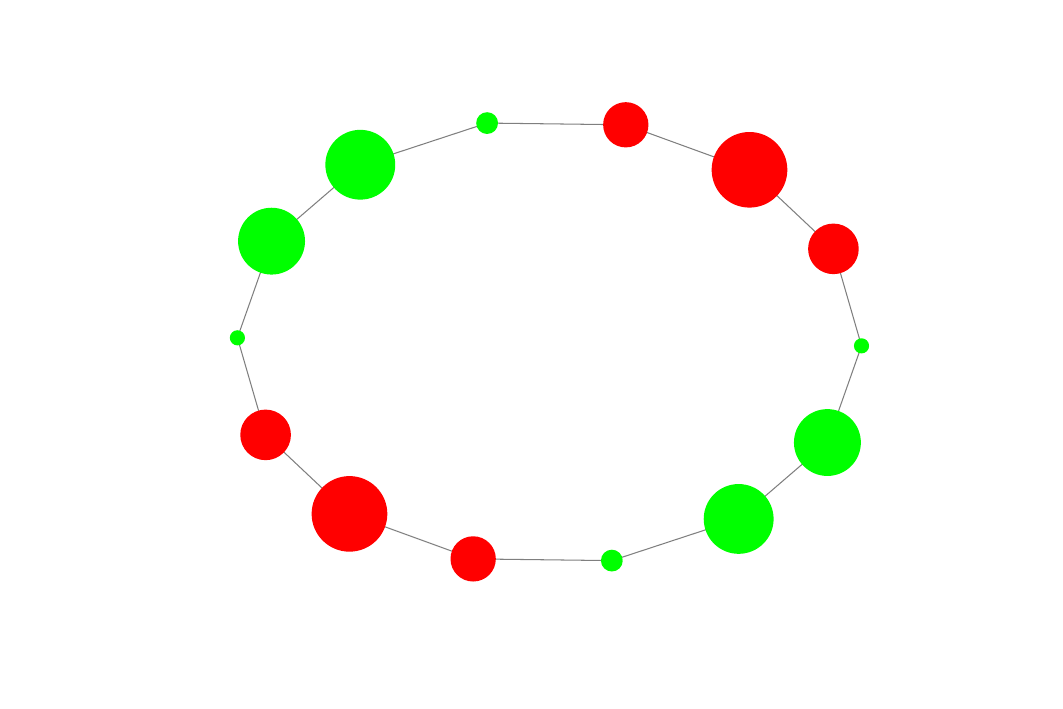}
        \;\; \includegraphics[scale=0.22]{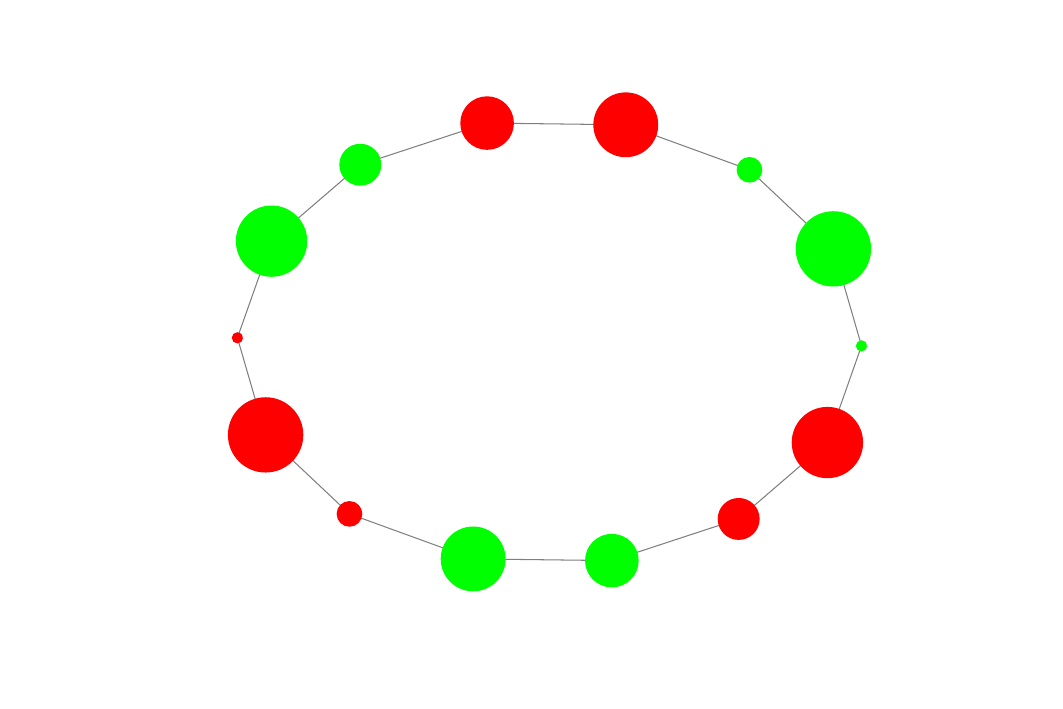}
		\\
		\includegraphics[scale=0.22]{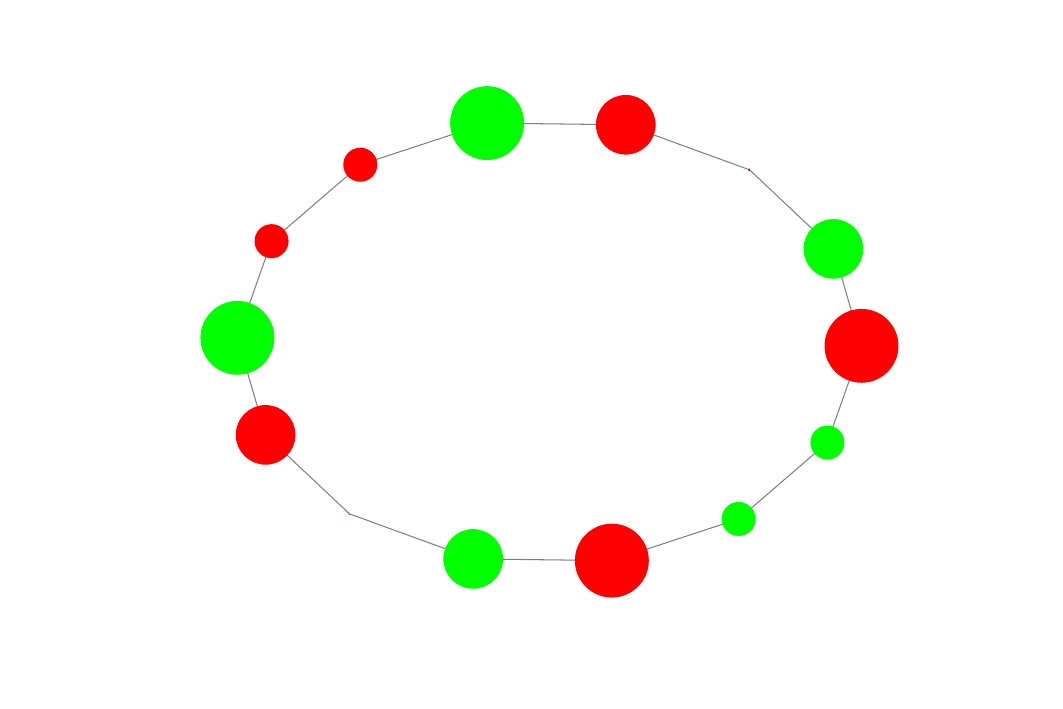}	
		\;\; \includegraphics[scale=0.22]{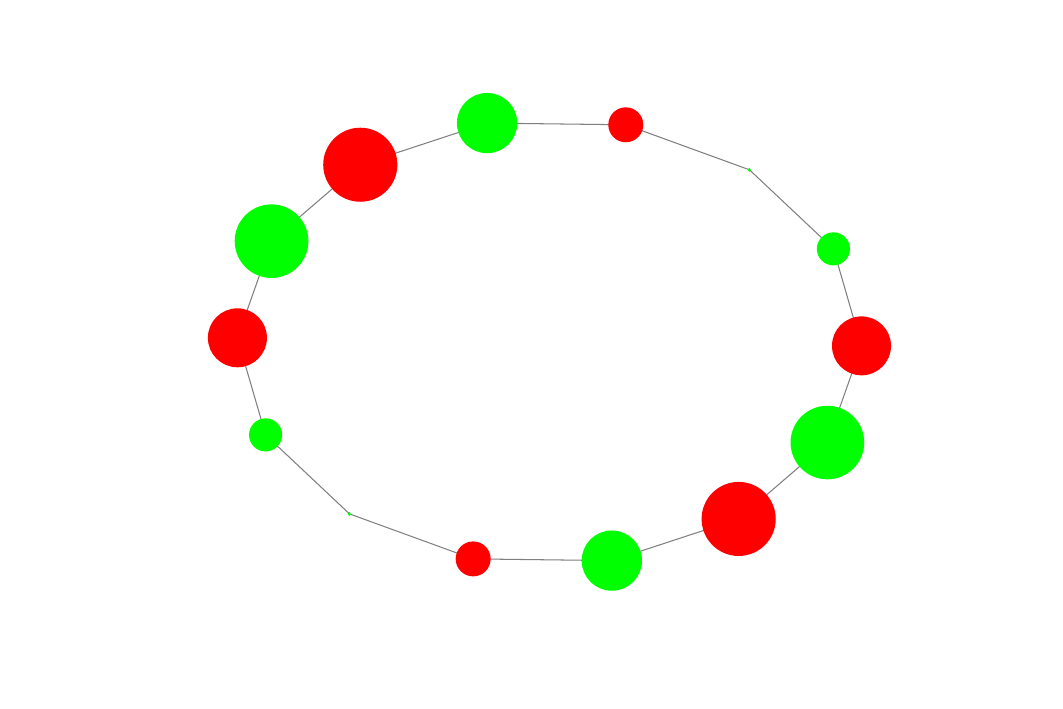}	
      \;\;\includegraphics[scale=0.22]{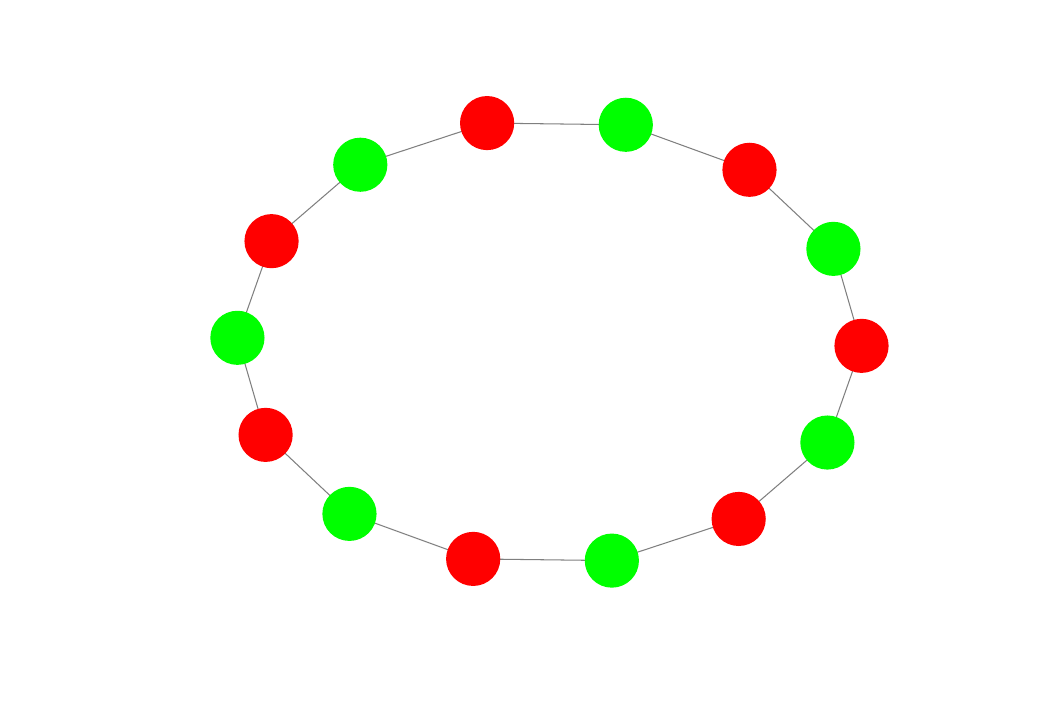}	
\caption{(Top) Eigenvectors 2, 4, 6. (Bottom) Eigenvectors 10, 12, 14. }
\label{Figure:principalcomponents}
\end{figure}

Figure \ref{Figure:principalcomponents} illustrates some eigenvectors/principal components of a circle network with 14 nodes and with links all having equal weight given by $1$. For each principal component, the color of a node indicates the sign of the entry of that node in that principal component (the color red means negative), while the size of a node indicates the absolute value of that entry.  A general feature that is worth noting is that the entries of the top principal components (smaller values of $\ell$) are clustered among neighboring nodes, while the bottom principal components (larger values of $\ell$)  tend to be negatively correlated among neighboring nodes.\footnote{The circle network is nongeneric in that eigenvectors are not uniquely determined, but an arbitrarily small perturbation of $\bm{G}$ will select a unique basis very close to the one depicted.}

\subsection{Analysis of the game using principal components} \label{sec:results}
For any vector $\bm{z} \in \mathbb{R}^n$, let $\underline{\bm{z}}=\bm{U}^\tr \bm{z}$. We will refer to $\underline{z}_\ell$ as the projection of $\bm{z}$ onto the $\ell^{\text{th}}$ principal component, or the magnitude of $\bm{z}$ in that component. Substituting the expression $\bm{G}=\bm{U} \bm{\Lambda} \bm{U}^\tr$ into equation (\ref{LS}),
which characterizes equilibrium, we obtain
\begin{equation*}
[\bm{I}-\beta \bm{U} \bm{\Lambda} \bm{U}^\tr] \bm{a}^*=\bm{b}.
\end{equation*}
Multiplying both sides of this equation by $\bm{U}^{\tr}$ gives us an analogue of (\ref{eq:LS-solved}) characterizing the solution of the game:
\begin{equation*}
		[\bm{I}-\beta \bm{\Lambda}]\underline{\bm{a}}^*=\underline{\bm{b}} \qquad \Longleftrightarrow \qquad  \underline{\bm{a}}^*=[\bm{I}-\beta \bm{\Lambda}]^{-1}\underline{\bm{b}}.
\end{equation*}
This system is diagonal, and the $\ell^{\text{th}}$ diagonal entry of $[\bm{I}-\beta \bm{\Lambda}]^{-1}$ is $\frac{1}{1-\beta\lambda_\ell}$. Hence, for every $\ell \in \{1,2,\ldots,n\}$,
		\begin{equation}\label{DecSolu}
		\underline{a}^*_\ell=\frac{1}{1-\beta\lambda_\ell}\underline{b}_\ell.
		\end{equation}
The principal components of $\bm{G}$ constitute a basis in which strategic effects are easily described. The equilibrium action $\underline{a}^*_\ell$ in the $\ell^{\text{th}}$ principal component of $\bm{G}$ is the product of an amplification factor (determined by the strategic parameter $\beta$ and the eigenvalue $\lambda_\ell$) and $\underline{b}_\ell$, which is simply the projection of $\bm{b}$ onto that principal component. Under Assumption \ref{as:spectral_radius}, for all $\ell$ we have $1-\beta\lambda_\ell > 0$.\footnote{Assumption \ref{as:spectral_radius} on the spectral radius implies that $\beta \bm{\Lambda}$ has no entries larger than $1$.} Moreover, when $\beta>0$ ($\beta<0$), the amplification factor is decreasing (increasing) in $\ell$.

We can also use this to give a formula for equilibrium actions in the original coordinates:
\begin{equation*}\label{BC-new}
a^*_i=\sum_{\ell}\frac{1}{1-\beta\lambda_\ell}u_{i}^\ell \underline{b}_\ell.
\end{equation*}


\medskip

We close with a definition that will allow us to describe optimal interventions in terms of a standard measure of their similarity to various principal components. 
\begin{definition}
The cosine similarity of two nonzero vectors $\bm{z}$ and $\bm{y}$ is
\[
\rho(\bm{z},\bm{y})=\frac{\bm{z}\cdot\bm{y}}{\Vert\bm{z}\Vert \Vert\bm{y}\Vert}.
\]
\end{definition}
\noindent This is the cosine of the angle between the two vectors in the plane determined by $\bm{y}$ and $\bm{z}$. When $\rho(\bm{z},\bm{y})=1$, vector $\bm{z}$ is a positive scaling of $\bm{y}$. When $\rho(\bm{z},\bm{y})=0$, vectors $\bm{z}$ and $\bm{y}$ are orthogonal. When $\rho(\bm{z},\bm{y})=-1$, vector $\bm{z}$ is a negative scaling of $\bm{y}$.

\section{Optimal interventions}\label{sec:optimal_targetS}

This section develops a characterization of optimal interventions and studies their properties.

We begin by dispensing with a straightforward case of the planner's problem. Recall that under Property A, the planner's payoff as a function of the equilibrium actions $\bm{a}^*$ is
$W(\bm{b},\bm{G})= w \cdot \left( \bm{a}^* \right)^{\tr} \bm{a}^*$. If $w<0$, the planner wishes to minimize the sum of the squares of the equilibrium actions. In this case, when the budget is large enough, that is, $C\geq \Vert\hat{\bm{b}}\Vert^2$, the planner can allocate resources to ensure that individuals have a zero target action by setting $b_i=0$ for all $i$. It follows from the best-response equations that all individuals choose action $0$ in equilibrium, and so the planner achieves the first-best.\footnote{In the local public good application (recall Example \ref{Ex4}) $w=-1$, and so when $C\geq \Vert\hat{\bm{b}}\Vert$, the optimal intervention satisfies $b^{*}_i=0$. Recalling our change of variables there ($b_i=[\tau-\tilde{b}]/2$), the optimal intervention in that case is to modify the endowment of each individual so that everyone accesses the optimal level of the local public good without investing personally.} The next assumption implies that the planner's bliss point cannot be achieved, so that there is an interesting optimization problem:
\begin{assumption}\label{Ass3}
Either $w<0$ and $C<\Vert\hat{\bm{b}}\Vert$, or $w>0$.
\end{assumption}

Let $\bm{b}^{*}$ solve the incentive-targeting problem (IT), and let $\bm{y}^*=\bm{b}^*-\hat{\bm{b}}$ be the vector of changes in individuals' standalone marginal benefits at the optimal intervention. Furthermore, let $$\alpha_\ell=\frac{1}{(1-\beta\lambda_\ell)^2}$$ and note that $a^*_{\ell}=\sqrt{\alpha_\ell}\underline{b}_\ell$ is the equilibrium action in the $\ell^{\text{th}}$ principal component of $\bm{G}$ (see equation (\ref{DecSolu})).

\begin{theorem}\label{Theorem1}
Suppose Assumptions  1--3 hold and the network game satisfies Property A. At the optimal intervention, the similarity between $\bm{y}^*$ and principal component $\bm{u}^\ell(\bm{G})$ satisfies the following proportionality:
\begin{eqnarray}\label{eq1}
\rho(\bm{y^*},\bm{u}^\ell(\bm{G})) \quad \propto \quad
\rho(\hat{\bm{b}},\bm{u}^\ell(\bm{G}))\frac{w\alpha_\ell}{\mu-w\alpha_\ell}, \quad \text{$\ell = 1,2,\ldots,n$,}
\end{eqnarray}
where $\mu$, the shadow price of the planner's budget, is uniquely determined as the solution to
\begin{equation}\label{LM}
\sum_{\ell}\left(\frac{w\alpha_{\ell}}{\mu-w\alpha_\ell}\right)^2\underline{\hat{b}}_{\ell}^{2}=C
\end{equation} and satisfies $\mu > w\alpha_\ell$ for all $\ell$, so that all denominators are positive.
\end{theorem}

We briefly sketch the main argument here and interpret the quantities in the formula. 
Define $x_{\ell}=(\underline{b}_{\ell}-\underline{\hat{b}}_{\ell})/\underline{\hat{b}}_{\ell}$ as the change of $\underline{b}_{\ell}$, relative to $\underline{\hat{b}}_\ell$. By rewriting the principal's objective and budget constraints in terms of principal components and plugging in the equilibrium condition (\ref{DecSolu}), we can rewrite the maximization problem  as
\[
\max_{\bm{x}} W(\bm{b},\bm{G})=\sum_{\ell} w\alpha_{\ell}(1+x_{\ell})^2\hat{\underline{b}}_\ell^2 \quad \text{s.t.} \quad \sum_{\ell}\underline{\hat{b}}^2_{\ell}x^{2}_{\ell} \leq C. 
\]
If the planner allocates a marginal unit of the budget to principal component $\ell$, the condition for equality of the marginal return  and marginal cost (recalling that $\mu$ is the multiplier on the budget constraint) is
\[
\underbrace{2\hat{\underline{b}}_\ell^2 \cdot w \alpha_\ell(1+x_\ell)}_{\text{marginal return}} \quad = \quad \underbrace{2\hat{\underline{b}}_\ell^2 \cdot \mu x_\ell}_{\text{marginal cost}}.
\]
It follows that $\frac{w \alpha_\ell}{\mu-w\alpha_\ell}$ is exactly the value of $x_\ell$ at which the marginal return  and the marginal cost are equalized.\footnote{It can be verified that the ratio for every $\ell\in\{1,\ldots,n-1\}$, $x_{\ell}/x_{\ell+1}$ is increasing (decreasing) in $\beta$ for the case of strategic complements (substitutes): thus the intensity of the strategic interaction shapes the relative importance of different principal components.} Rewriting $x_\ell$ in terms of cosine similarity, that equality implies $$\frac{w \alpha_\ell}{\mu-w\alpha_\ell}= x_\ell^* = \frac{\Vert \bm{y}^* \Vert \text{ } \rho(\bm{y}^*,\bm{u}^\ell(\bm{G}))}{\Vert \hat{\bm{b}} \Vert \text{ } \rho(\hat{\bm{b}},\bm{u}^\ell(\bm{G}))}.$$ Rearranging this yields the proportionality expression (\ref{eq1}) in the theorem. The Langrange multiplier $\mu$ is determined by solving the simple equation (\ref{LM}). Now, given $\mu$,  the similarities $\rho(\bm{y}^*,\bm{u}^\ell(\bm{G}))$ determine the direction of the optimal intervention
$\bm{y}^*$. The magnitude of the intervention is found by exhausting the budget.
Thus Theorem \ref{Theorem1} provides a full characterization of the optimal intervention.

Next, we discuss the formula for the similarities given in expression (\ref{eq1}). The similarity between $\bm{y}^*$ and $\bm{u}^\ell(\bm{G})$ measures the extent to which principal component $\bm{u}^\ell(\bm{G})$ is represented in the optimal intervention $\bm{y}^*$. Equation (\ref{eq1}) tells us that this is proportional to two factors. The first factor, $\rho(\hat{\bm{b}},\bm{u}^\ell(\bm{G}))$, is a status quo effect corresponding to the similarity between the $\ell^{\text{th}}$ principal component and the status-quo vector $\hat{\bm{b}}$. This factor summarizes how much the initial condition influences the optimal intervention for a given budget. The intuition here is that if a given principal component is strongly represented in the status quo vector of standalone incentives, then -- because of the convexity of welfare in the principal component basis -- changes in that dimension have a particularly significant effect.

The second factor, $\frac{w\alpha_\ell}{\mu-w\alpha_\ell}$, is determined by two quantities: the eigenvalue corresponding to $\bm{u}^\ell(\bm{G})$  (via $\alpha_\ell = \frac{1}{1-\beta \lambda_\ell}$), and the budget $C$ (via  the shadow price $\mu$).  To focus on this second factor, $\frac{w\alpha_\ell}{\mu-w\alpha_\ell}$, we define the \emph{similarity ratio} of $\bm{u}^\ell(\bm{G})$ to be the fraction
\[
r_\ell^*=\frac{\rho(\bm{y^*},\bm{u}^\ell(\bm{G}))}{\rho(\hat{\bm{b}},\bm{u}^\ell(\bm{G}))}.
\]
Theorem \ref{Theorem1} shows that, as we vary $\ell$, the similarity ratio $r_\ell^*$ is proportional to $\frac{w \alpha_\ell}{\mu-w\alpha_\ell}$. It follows that the similarity ratio is greater, in absolute value, for the principal components $\ell$ with greater $\alpha_\ell$. Intuitively, those are the components where the intervention makes the largest change relative to the status quo profile of incentives. The ordering of these coefficients corresponds to the eigenvalues in a way that depends on the nature of strategic spillovers:
\begin{corollary}\label{cr:monotonicity}
Suppose Assumptions 1--3 hold and the network game satisfies Property A. If the game is one of strategic complements ($\beta>0$), then $|r_\ell^*|$ is decreasing in $\ell$; if the game is one of strategic substitutes ($\beta<0$), then $|r_\ell^*|$ is increasing in $\ell$.
\end{corollary}

In some problems there may be a nonnegativity constraint on actions, in addition to the constraints in problem  (IT). Note that as long as the status quo actions $\hat{\bm{b}}$ are positive, this constraint will be respected for all $C$ less than some $\hat{C}$, and so our approach will give information about the relative effects on various components for interventions that are not too large.

\subsection{Small and large budgets}
The optimal intervention takes especially simple forms in the cases of small and large budgets. From equation
(\ref{LM}), we can deduce that the shadow price $\mu$ is decreasing in $C$. For $w>0$, it follows that an increase in $C$ raises
$\frac{w \alpha_\ell}{\mu-w\alpha_\ell}$ and that the principal components with larger $\alpha_\ell$ become larger in relative terms as well; in other words, if $w>0$ and $\alpha_\ell>\alpha_{\ell'}$, then $r^*_\ell/r^*_{\ell'}$ is increasing in $C$.\footnote{Analogously, when $w<0$, $\frac{w \alpha_\ell}{\mu-w\alpha_\ell}$ and $r^*_\ell/r^*_{\ell'}$ are both decreasing in $C$.} For simplicity of exposition, we suppress the dependence of outcomes on $C$ in the following statement, but note that $\bm{y}^*$ and thus the $r_\ell^*$ are all functions of $C$.

\begin{proposition}\label{result:smallC}
Suppose Assumptions 1--3 hold and the network game satisfies Property A. Then the following hold:
\begin{itemize}
\item[1.] As $C \to0$, in the optimal intervention, $\frac{r_\ell^*}{r_{\ell'}^*}\to  \frac{\alpha_\ell}{\alpha_{\ell'}}$.
\item[2.] As $C \to \infty$, in the optimal intervention
\begin{itemize}
\item[2a.] If the game has the strategic complements property, $\beta>0$, then the similarity of $\bm{y}^*$ and the first principal component of the network tends to $1$, $\rho(\bm{y^*},\bm{u}^1(\bm{G}))\to 1$.
\item[2b.] If the game has the strategic substitutes property, $\beta<0$, then the similarity of $\bm{y}^*$ and the last principal component of the network tends to $1$, $\rho(\bm{y^*},\bm{u}^n(\bm{G}))\to1$.
\end{itemize}
\end{itemize}
\end{proposition}

This result can be understood by recalling equation (\ref{eq1}) in Theorem \ref{Theorem1}. First, consider the case of small $C$. When the planner's budget becomes small, the shadow price $\mu$ tends to $\infty$.\footnote{As costs are quadratic, small relaxation in the budget around zero can have a large impact on aggregate welfare.} Equation (\ref{eq1}) then implies that the similarity ratio of the $\ell^{\text{th}}$ principal component becomes proportional to $\alpha_\ell$.  Turning now to the case where $C$ grows large, the shadow price converges to $w \alpha_1$ if $\beta>0$, and to $w \alpha_n$ if $\beta<0$ (by equation (\ref{LM})). Plugging this into equation (\ref{eq1}), we find that in the case of strategic complements, the optimal intervention shifts individuals' standalone marginal returns (very nearly) in proportion to the first principal component of $\bm{G}$, so that $\bm{y}^*\to \sqrt{C} \bm{u}^1(\bm{G})$. In the case of strategic substitutes, on the other hand, the planner changes individuals' standalone marginal returns (very nearly) in proportion to the last principal component, namely $\bm{y}^*\to \sqrt{C}\bm{u}^n(\bm{G})$.\footnote{When individuals' initial standalone marginal returns are zero ($\hat{\bm{b}}=\bm{0}$), we can dispense with the approximations invoked for a large budget $C$. Assuming that $\bm{G}$ is generic, if  $\hat{\bm{b}}=\bm{0}$ then, regardless of the level of $C$, the entire budget is spent either (i) on changing $\underline{b}_1$ (if $\beta >0$) or (ii) on changing $\underline{b}_n$ (if $\beta<0$). To see this, consider the proof of Theorem \ref{Theorem1} and set $\hat{\underline{\bm{b}}}=\bm{0}$ in the maximization problem (IT-PC), the principal component version of (IT). Note that if the allocation is not extreme, then the effort can be reallocated profitably among the principal components without changing the cost.}

\begin{figure}
\centering
  \begin{subfigure}{.48\textwidth}
\centering
 \includegraphics[width=1\linewidth]{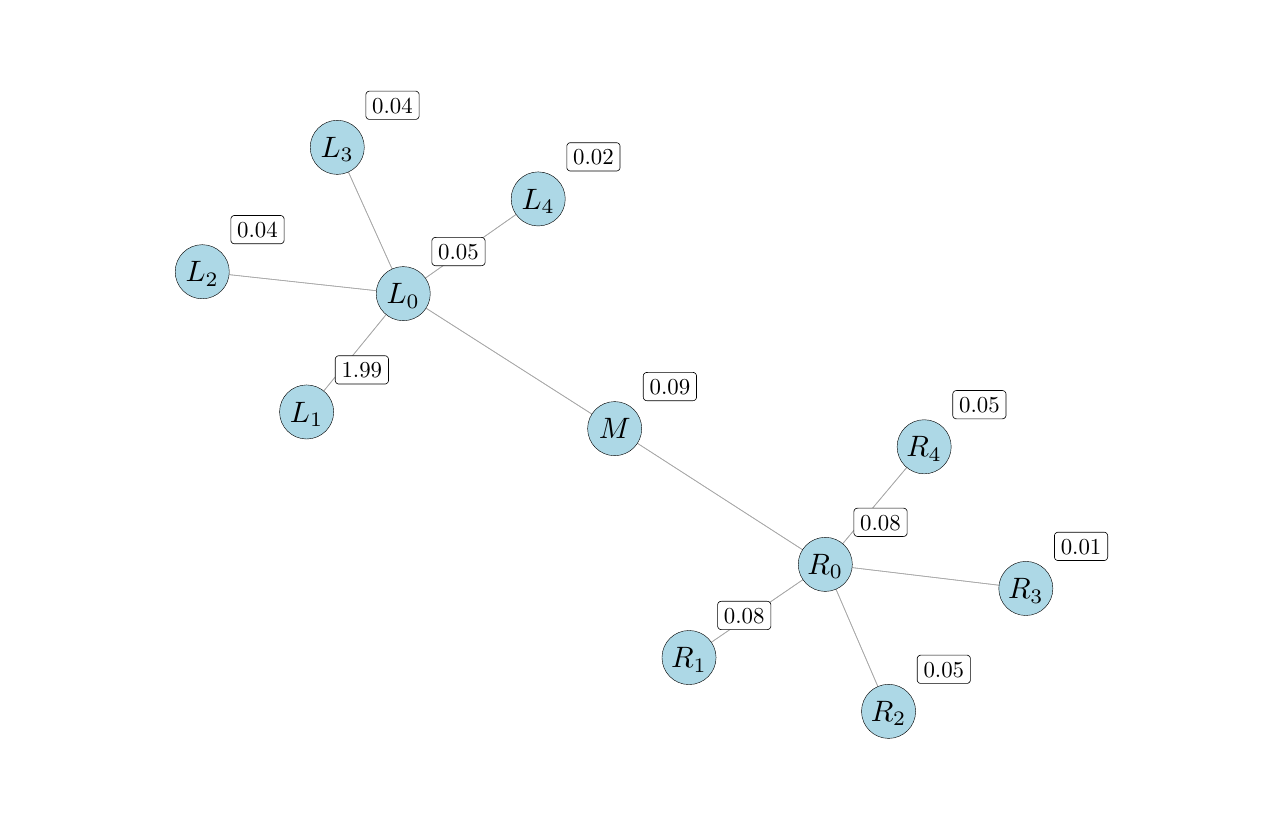}
\caption{\scriptsize  The network}
\label{fig:simplegraph}
\end{subfigure}
   \begin{subfigure}{.48\textwidth}
 \centering
    \includegraphics[width=1\linewidth]{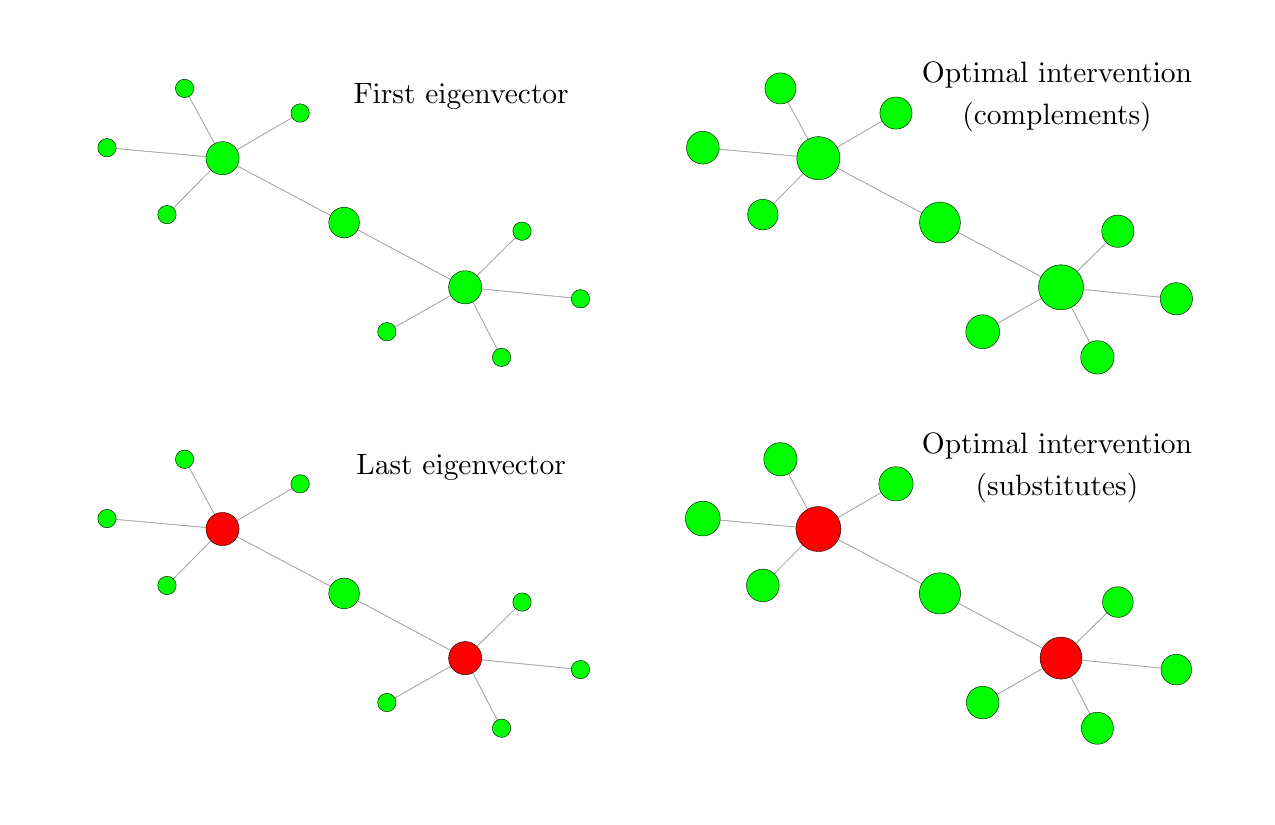}
    \caption{\scriptsize  Eigenvectors and Interventions}
        \end{subfigure}
\caption{Optimal targets with large budgets}\label{fig2ii}
\end{figure}

Figure \ref{fig2ii} presents optimal targets when the budget is large -- in particular, for $C=500$. We consider an 11-node undirected network with binary links containing two hubs, $L_0$ and $R_0$, that are connected by an intermediate node $M$; the network is depicted in Figure 2(A). The numbers next to the nodes are the status quo standalone marginal returns. Payoffs are as in Example \ref{Ex1}. For the case of strategic complements, we set $\beta=0.1$, and for strategic substitutes we set $\beta=-0.1$. Assumptions \ref{symmetry} and \ref{as:spectral_radius} are satisfied and Property A holds. The top-left of Figure 2(B) illustrates the first eigenvector, and the top-right depicts optimal targets in a game with strategic complements. The bottom-left of Figure 2(B) illustrates the last eigenvector, and the bottom-right depicts the optimal targets when the game has strategic substitutes. The node size represents the size of the intervention, $|b^*_i-\hat{b}_i|$; its color represents the sign of the intervention (with green signifying a positive intervention and red indicating a negative intervention).

In line with part 2 of Proposition \ref{result:smallC}, for large $C$, the optimal intervention is guided by the ``main'' component of the network (corresponding to the largest or smallest eigenvalue).  Under strategic complements, this is the first eigenvector of the network, which corresponds to individuals' eigenvector centrality.\footnote{Online Appendix Section \ref{OA-sec:othermeasures} presents a discussion of eigenvector centrality.} Intuitively, by increasing the standalone marginal return of each individual in proportion to his eigenvector centrality, the planner targets the individuals in proportion to their global contributions to strategic feedbacks, and this is welfare maximizing.

Under strategic substitutes, optimal targeting is determined by the last eigenvector of the network, corresponding to its smallest eigenvalue. This network component contains information about the local structure of the network: it  determines the way to partition the set of nodes into two sets so that most of the links are across individuals in different sets.\footnote{The last eigenvector of a graph is useful in determining the bipartiteness of a graph and its chromatic number. \citet{desai1994characterization} characterize the smallest eigenvalue of a graph and relate it to the degree of bipartiteness of a graph. \citet{alon1997spectral} relate the last eigenvector  to a coloring of the underlying graph, that is, a labeling of nodes by a minimal set of integers such that no neighboring nodes share the same label.} The optimal intervention increases the standalone marginal returns of all individuals in one set and decreases those of individuals in the other set. The planner wishes to target neighboring nodes asymmetrically, as this reduces possible crowding-out effects that occur due to the strategic substitutes property.



\subsection{When are interventions simple?} \label{sec:simple}

We have just seen examples illustrating how, with large budgets, the intervention is simple: approximately proportional to just one principal component -- the top or bottom one. After defining simplicity formally, our final result in this section characterizes how large the budget must be for this approximation to be a close one.

\begin{definition}[Simple interventions]
An intervention is \emph{simple} if, for all $i\in \mathcal{N}$,
\begin{itemize}
\item $b_i-\hat{b}_{i}=\sqrt{C}u^{1}_{i}$ when the game has the strategic complements property ($\beta>0$),
\item $b_i-\hat{b}_{i}=\sqrt{C}u^{n}_{i}$ when the game has the strategic substitutes property ($\beta<0$).
\end{itemize}
Let $W^*$ be the aggregate utility under the optimal intervention, and let $W^s$ be the aggregate utility under the simple intervention.
\end{definition}

\begin{proposition}\label{Prop:LargeBudget}
Suppose $w>0$, Assumptions 1 and 2 hold, and the network game satisfies Property A. 
\begin{itemize}
\item[1.] If the game has the strategic complements property, $\beta>0$, then for any $\epsilon>0$, if $C>\frac{2\Vert\hat{\bm{b}}\Vert^2}{\epsilon}\left(\frac{\alpha_2}{\alpha_1-\alpha_2}\right)^2$, then $W^*/W^s<1+\epsilon$ and $\rho(\bm{y}^*,\sqrt{C}\bm{u}^1)>\sqrt{1-\epsilon}$.
\item[2.] If the game has the strategic substitutes property, $\beta<0$, then for any $\epsilon>0$, if $C>\frac{2\Vert\hat{\bm{b}}\Vert^2}{\epsilon}\left(\frac{\alpha_{n-1}}{\alpha_n-\alpha_{n-1}}\right)^2$, then $W^*/W^s<1+\epsilon$ and $\rho(\bm{y}^*,\sqrt{C}\bm{u}^n)>\sqrt{1-\epsilon}$.
\end{itemize}
\end{proposition}
Proposition \ref{Prop:LargeBudget} gives a condition on the size of the budget beyond which (a) simple interventions achieve most of the optimal welfare and (b) the optimal intervention is very similar to the simple intervention. This bound depends on the status quo standalone marginal returns and the structure of the network.

We first discuss the dependence on the status quo benefits. Observe that the first term on the right-hand side of the inequality for $C$ is proportional to ${\Vert\hat{\bm{b}}\Vert}$. This inequality is therefore easier to satisfy when the status quo standalone  marginal returns are smaller, in the sense of having a smaller norm. The inequality is harder to satisfy when these marginal returns are large and/or heterogeneous.\footnote{Recall that $ \Vert \frac{1}{n}\hat{\bm{b}} \Vert^2$ is equal to the sum of $\left(\frac{1}{n} \sum_i \hat{b}_i\right)^2$ (the squared mean of the entries of $\bm{b}$) and the sum of squared deviations of the entries of the vector $\hat{\bm{b}}$ from their mean.}

Next, consider the role of the network.  Recall that $\alpha_\ell = (1-\beta\lambda_\ell)^{-2}$; thus if $\beta>0$, the term $\alpha_2/(\alpha_1-\alpha_2)$ of the inequality is large when $\lambda_1-\lambda_2$, the ``spectral gap'' of the graph, is small. 
If $\beta<0$, then the term $\alpha_{n-1}/(\alpha_{n-1}-\alpha_n)$ is small when the ``bottom gap" of the graph, the difference $\lambda_{n-1}-\lambda_n$, is small. 

\begin{figure}
    \begin{subfigure}{.48\textwidth}
    \centering
           \includegraphics[width=1\linewidth]{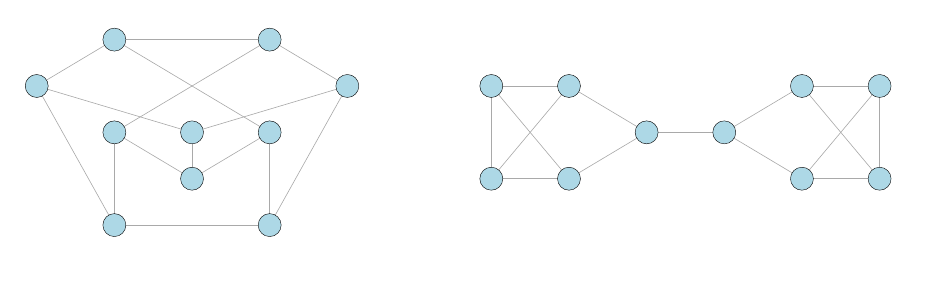}
\caption{\scriptsize Large (left) and small (right) spectral gap}
\label{fig:spectralgap}
       \end{subfigure}
    ~ 
      \quad
    \begin{subfigure}{.48\textwidth}
      \centering
            \includegraphics[width=1\linewidth]{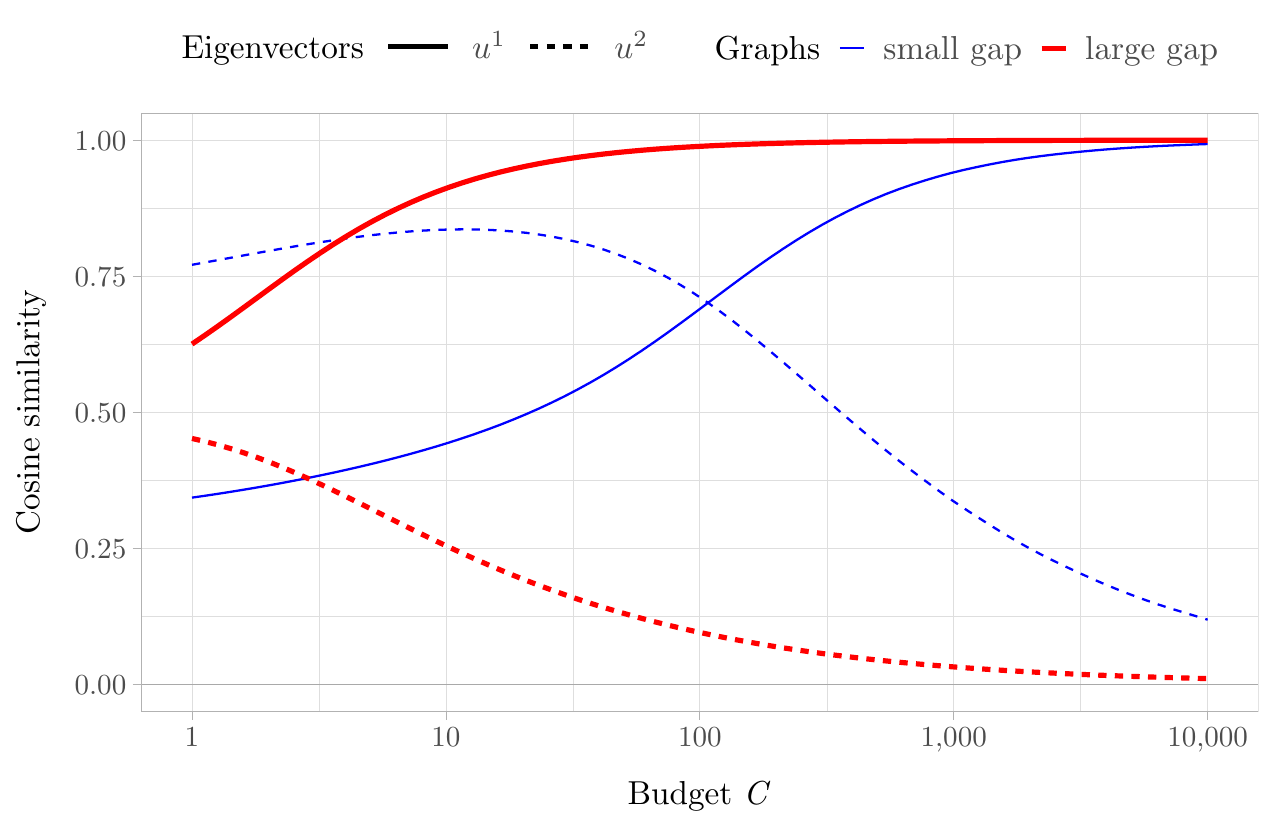}
        \caption{\scriptsize Cosine similarity and strategic complements}
        \label{fig:cosinenetworkSG}
    \end{subfigure}
    \hfill

    ~ 

    \begin{subfigure}{.48\textwidth}
    \centering
           \includegraphics[width=1\linewidth]{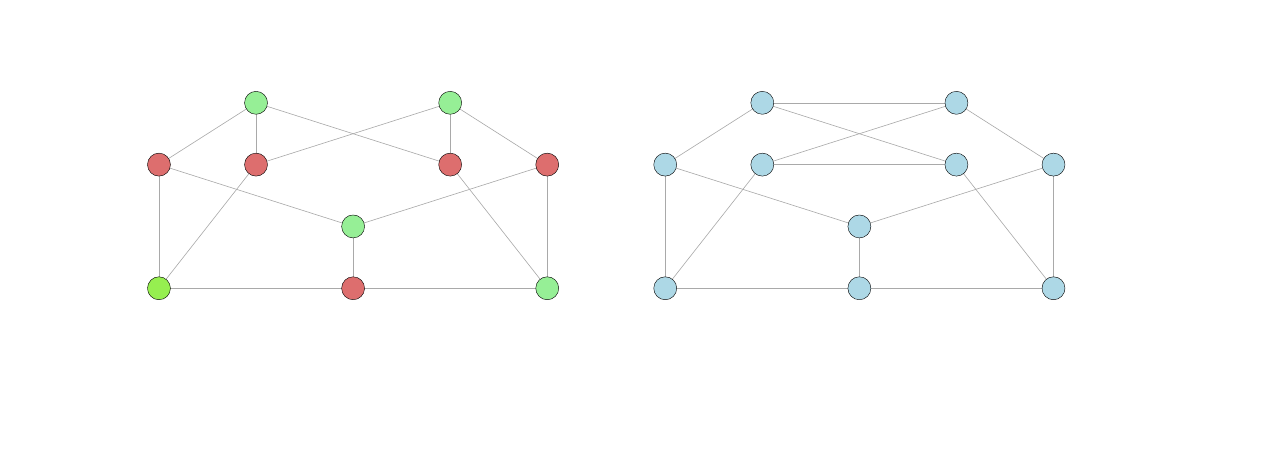}
\caption{\scriptsize Large (left) and small (right) bottom gap}
\label{fig:bottomgaps}
       \end{subfigure}
       \quad
         \begin{subfigure}{.48\textwidth}
         \centering
        \includegraphics[width=1\linewidth]{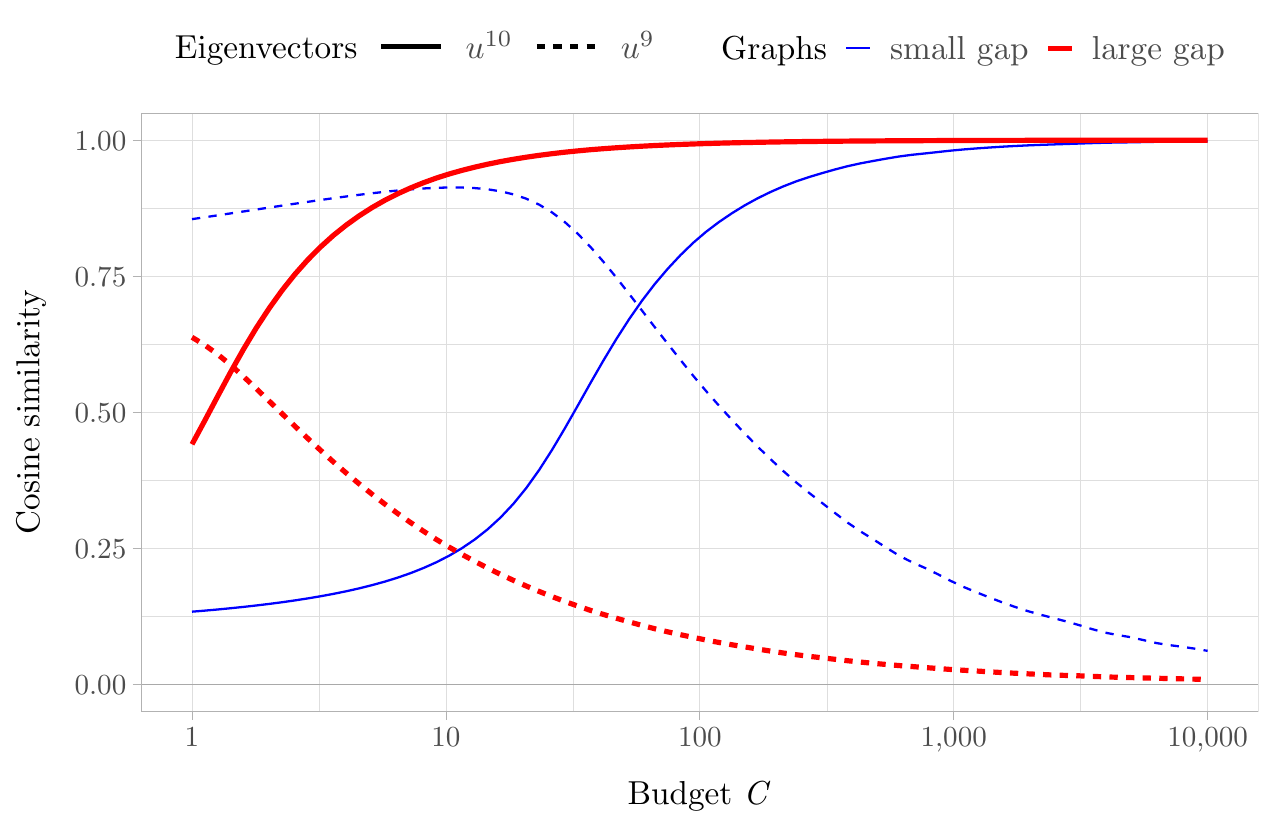}
        \caption{\scriptsize Cosine similarity and strategic substitutes}
        \label{fig:cosinenetworkBG}
    \end{subfigure}
    \caption{Spectral gap, bottom gap, and optimal interventions}\label{fig2}
\end{figure}

We now examine what network features affect these gaps, and illustrate  with examples, depicted in Figure \ref{fig2}. The obstacle to simplicity is a strong dependence on the status quo standalone marginal benefits. This dependence will be strong when two different principal components in the network offer similar amplification (all else equal) of interventions in that component. Which of these principal components receives the planner's focus will depend strongly on the status quo. In such networks, interventions will \emph{not} be simple for reasonable budgets. The merit of Proposition \ref{Prop:LargeBudget} is to show that small spectral or small bottom gap capture this property of the network. Figure \ref{fig2} illustrates the role of the network structure in shaping the rate (in terms of the size of the budget $C$) at which the optimal intervention converges to a simple intervention as we vary C. Under strategic complements, the optimal intervention converges to a simple one faster (as we vary $C$) in a network that has a large spectral gap. Under strategic substitutes, the optimal intervention converges to a simple one faster (as we vary $C$) in a network that has a large bottom gap.

We now describe which network properties, at a more intuitive level, correspond to having small and large spectral gaps. First, consider the case of strategic complements. A standard fact is that the two largest eigenvalues can be expressed in terms of the corresponding eigenvectors as follows:
\[
\lambda_1=\max_{\bm{u}:\Vert\bm{u}\Vert=1}\sum_{ij}g_{ij}u_i u_j \qquad \lambda_{2}=\max_{\substack{\bm{u} \colon \Vert \bm{u} \Vert = 1 \\ \bm{u} \cdot \bm{u}^1=0} }\sum_{ij}g_{ij}u_i u_j.
\]
Eigenvector $\bm{u}^1=\arg\max_{\bm{u}:\Vert\bm{u}\Vert=1}\sum_{ij}g_{ij}u_i u_j $ (corresponding to $\lambda_1$) assigns the same sign -- say, positive to all nodes in the network. Clearly, eigenvector $\bm{u}^2$ must assign negative values to some of the nodes (as it is orthogonal to $\bm{u}^1$). In the network on the left side of Figure 4(A) any such assignment will result in many adjacent nodes having opposite sign entries of $\bm{u}^2$; as a result, many terms in the expression for $\lambda_2$ will be negative, and $\lambda_2$ will be much smaller than $\lambda_1$, leading to a large spectral gap. In the network on the right side of  4(A), $\bm{u}^2$ will have positive-sign entries to nodes in one community and negative-sign entries for nodes in the other community. Because there are few edges between the communities, $\lambda_2$ turns out to be almost as large as $\lambda_1$. This will yield a small spectral gap. Thus, spectral gap measures the level of ``cohesiveness'' of the network, and it is this property that dictates fast convergence to simple interventions.\footnote{See  \cite{hartfiel1998structure}, \cite{levin-peres-wilmer}, and \cite{golub-jackson-homophily} for discussions and further citations to the literature on spectral gaps.}

Turning next to strategic substitutes, recall that the smallest two eigenvalues, $\lambda_n$ and $\lambda_{n-1}$, can be written in terms of the corresponding eigenvectors as follows:
\begin{equation}
\lambda_n=\min_{\bm{u}:\Vert\bm{u}\Vert=1}\sum_{ij}g_{ij}u_i u_j \qquad \lambda_{n-1}=\min_{\substack{\bm{u} \colon \Vert \bm{u} \Vert = 1 \\ \bm{u} \cdot \bm{u}^n=0} }\sum_{ij}g_{ij}u_i u_j. \label{eq:lambda_bottom}
\end{equation}
This tells us that $|\lambda_n|$ is large when the eigenvector $\bm{u}^n=\arg\min_{\bm{u}:\Vert\bm{u}\Vert=1}\sum_{ij}g_{ij}u_i u_j $ (corresponding to $\lambda_n$) assigns opposite signs to most pairs of adjacent nodes. In other words, the last eigenvalue is large when nodes can be partitioned into two sets and most of the connections are across sets, and thus $|\lambda_n|$ is maximized in a bipartite graph. The second-smallest eigenvalue of $\bm{G}$ reflects the extent to which the next-best eigenvector (orthogonal to $\bm{u}^n$) is good at solving the same minimization problem. Hence, the bottom gap of $\bm{G}$ is small when there are two orthogonal ways to partition the network into two sets so that, either way, the ``quality'' of the bipartition, as measured by $\sum_{ij}g_{ij}u_i u_j$, is similar.

We illustrate Proposition \ref{Prop:LargeBudget} with a comparison of the two graphs in Figure 4(C). The left-hand graph is bipartite: the last eigenvalue is $\lambda_n=-3$ and the second last eigenvalue is $\lambda_{n-1}=-1.64$.  By contrast, in the graph on the right of Figure 4(C), the bottom eigenvalue $\lambda_n=-2.62$, while the second lowest is $\lambda_{n-1}=-2.30$. This yields a much smaller bottom gap.\footnote{Intuitively, because $\bm{u}^n$ does not correspond to a perfect bipartition, it is easier for a vector orthogonal to  $\bm{u}^n$ to achieve a similarly low value of $\sum_{ij}g_{ij}u_i u_j$. }
This difference in bottom gap is reflected in optimal targeting policy in Figure 4(D): in the graph with large bottom gap, the optimal intervention puts little weight on the eigenvector $\bm{u}^{n-1}$  for a relatively small budget; it takes a much larger budget under a small bottom gap.

We conclude by noting the influence of the status quo standalone marginal returns in shaping optimal interventions for small budgets. For a small budget $C$, the cosine similarity of the optimal intervention for non-main network components can be higher than the one for the main component. This is true when the status quo $\hat{\bm{b}}$ is similar to some of the non-main network components; see Figures 4(B) and Figure 4(D).


\vspace{-.1in}
\section{Incomplete information}\label{sec:incompleteinformation}

In the basic model, we assumed that the planner knows the standalone marginal returns of every individual. This section extends the analysis to settings where the planner does \emph{not} know these parameters. 
For ease of exposition, we focus on network games that satisfy Property A.

Formally, fix a probability space $(\Omega, \mathcal{F}, \mathbb{P})$. The planner's belief over states is given by $\mathbb{P}$. This represents the planner's uncertainty, given all her information. The planner has control over the random vector (r.v.) $\mathcal{B}$, that is, a function $\mathcal{B}:\Omega\to\mathbb{R}^n$. The choice of $\mathcal{B}$ determines the cost of intervention. A realization of the random vector is denoted by $\bm{b}$.  This realization is common knowledge among individuals when they choose their actions. Thus, the game individuals play is one of complete information.\footnote{It is possible to go further and allow for incomplete information among the individuals about each other's $b_i$. We do not pursue this substantial generalization here; see \citet{GolubMorris2017-ENC} and \citet{Lambert2018NBER} for analyses in this direction.} We also define a function $K$ that gives the cost $K(\mathcal{B})$ of implementing the random variable $\mathcal{B}$.\footnote{The domain of this function is the set of all random vectors taking values in $\mathbb{R}^n$ defined on our probability space.}

We solve the following incomplete-information intervention problem:
\begin{align}
	\text{choose r.v. } \mathcal{B} \text{ to maximize }  & \mathbb{E} \left[ W(\bm{b};\bm{G}) \right] \tag{IT-G} \\
	\text{ s.t. } & [\bm{I}-\beta\bm{G}]\bm{a}^{*}=\bm{b}, \nonumber \\
& K(\mathcal{B}) \leq C . \nonumber
\end{align}
\noindent Note that the intervention problem (IT) under complete information is a special case of a degenerate r.v. $\mathcal{B}$: one in which the planner knows the vector of standalone marginal returns exactly and implements a deterministic adjustment relative to it.

To guide our modeling of the cost of intervention, we now review the features of the distribution of ${\mathcal{B}}$ that matter for aggregate welfare. For network games that satisfy Property A, we can write:
\begin{align}\mathbb{E}\left[ W(\bm{b};\bm{G}) \right] &= w\mathbb{E}[\left( \bm{a}^* \right)^{\tr} \bm{a}^*]=w\mathbb{E} [\underline{\bm{a}}^{\tr}\underline{\bm{a}}] =w\sum_\ell \alpha_\ell \left( \mathbb{E}[\underline{b}_\ell]^2+ \text{Var}[\underline{b}_\ell]\right). \label{eqn:W_var_formula}
\end{align}
In words, welfare is determined by the mean and variance of the realized components $\underline{b}_\ell$; these in turn are determined by the first and second moments of the chosen random variable $\mathcal{B}$. In view of this, we will consider intervention problems where the planner can modify the mean and the covariance matrix of $\mathcal{B}$, and the cost of intervention depends only on these modifications.

\subsection{Mean shifts}
We first consider an intervention where there is an arbitrarily distributed vector of standalone marginal returns and the planner's intervention shifts it in a deterministic way. Formally, fix a random variable $\hat{\mathcal{B}}$, called the status quo, with typical realization $\hat{\bm{b}}$. The planner's policy is given by $\bm{b}=\hat{\bm{b}}+\bm{y}$, where $\bm{y} \in \mathbb{R}^n$ is a deterministic vector. We denote the corresponding random variable by $\mathcal{B}_{\bm{y}}$. In terms of interpretation, note that implementing this policy does not require knowing $\hat{\bm{b}}$ as long as the planner has an instrument that shifts incentives.
\begin{assumption}\label{K1}
The cost of implementing r.v. $\mathcal{B}_{\bm{y}}$ is
\begin{equation*}
K(\mathcal{B}_{\bm{y}}) =\sum_i y_i^2,
\end{equation*}
and $K(\mathcal{B})$ is $\infty$ for any other random variable.
\end{assumption}
\noindent In contrast to the analysis of Theorem \ref{Theorem1}, the vector $\hat{\bm{b}}$ is a random variable. But we take the analogue of the cost function used there, noting that in the deterministic setting this formula held with $\bm{y} = \bm{b} - \hat{\bm{b}}$.

\begin{proposition}\label{Prop:IF1}
Consider problem (IT-G) with the cost of intervention satisfying Assumption \ref{K1}. Suppose Assumptions 1 and 2 hold and the network game satisfies Property A.
The optimal intervention policy $\mathcal{B}^*$ is equal to $\mathcal{B}_{\bm{y}^*}$, where $\bm{y}^*$ is the optimal intervention in the deterministic problem with $\overline{\bm{b}}= \mathbb{E} [\hat{\bm{b}}] $ as the status quo vector of standalone marginal returns.
\end{proposition}


\subsection{Intervention on variances}

We next consider the case where the planner faces a vector of means, fixed at $\bar{\bm{b}}$, and, subject to that, can choose any random variable $\mathcal{B}$. The difference in the expected welfare for two different interventions $\mathcal{B}$ and $\widetilde{\mathcal{B}}$ depends only on the variance--covariance matrix of $\mathcal{B}$ and $\widetilde{\mathcal{B}}$. Thus, the planner effectively faces \emph{the problem of intervening on variances}. We prove a result on optimal intervention for all cost functions satisfying certain symmetries.
\begin{assumption}\label{as:K-rotation}
The cost function satisfies two properties: (a) $K(\mathcal{B})=\infty$ if $\mathbb{E} \bm{b}\neq\bar{\bm{b}}$; (b) $K(\mathcal{B})=K(\widetilde{\mathcal{B}})$ if $\widetilde{\bm{b}}-\bar{\bm{b}}=  \bm{O}(\bm{b}-\bar{\bm{b}})$, where $\bm{O}$ is an orthogonal matrix. Analogous to our other notation, we use $\widetilde{\bm{b}}$ for realizations of the random vector with distribution $\widetilde{\mathcal{B}}$.
\end{assumption}
Part (a) is a restriction on feasible interventions, namely a restriction to interventions that are mean--neutral. Part (b) means that rotations of coordinates around the mean do not affect the cost of implementing a given distribution. This assumption gives the cost a directional neutrality, which ensures that our results are driven by the benefits side rather than by asymmetries operating through the costs. For example, let $\bm{\Sigma}_{\mathcal{B}}$ be the variance--covariance matrix of the random variable $\mathcal{B}$. In particular, $\sigma^{\mathcal{B}}_{ii}$ is the variance of $b_i$. Suppose that the cost of implementing $\mathcal{B}$ with $\mathbb{E} \bm{b}=\bar{\bm{b}}$ is a function of the sum of the variances of the $b_i$:
\begin{equation}\label{K2}
K(\mathcal{B}) = \left\{
  \begin{array}{lr}
    \phi\left(\sum_i \sigma^{\mathcal{B}}_{ii}\right) & \text{ if } \mathbb{E} \bm{b}=\bar{\bm{b}}\\
    \infty & \text{otherwise}.
  \end{array}
\right.
\end{equation}
The cost function ($\ref{K2}$) satisfies property (a) of Assumption \ref{as:K-rotation}. Moreover, it satisfies property (b) of Assumption \ref{as:K-rotation} because $\sum_i \sigma^{\mathcal{B}}_{ii}=\operatorname{trace}\bm{\Sigma}_{\mathcal{B}}$; this trace is the sum of the eigenvalues of $\bm{\Sigma}_{\mathcal{B}}$, which is invariant to the transformation defined in (b).\footnote{When we look at the variance--covariance matrix of  $\widetilde{\bm{b}}$ defined by $\widetilde{\bm{b}}-\bar{\bm{b}}=  \bm{O}(\bm{b}-\bar{\bm{b}})$, the variance--covariance matrix becomes $\bm{O} \bm{\Sigma} \bm{O}^\tr$, and this has the same eigenvalues and therefore the same trace.}

\begin{proposition}[Variance control]\label{Pr:AV}
Consider problem (IT-G) with the cost of intervention satisfying Assumption \ref{as:K-rotation}. Suppose Assumptions 1 and 2 hold and the network game satisfies Property A.
Let the optimal intervention be $\mathcal{B}^*$. We have the following:
	\begin{itemize}
		\item[1.] Suppose the planner likes variance (i.e., $w>0$). If the game has strategic complements ($\beta>0$), then $\operatorname{Var}( \bm{u}^\ell(\bm{G}) \cdot \bm{b}^*)$ is weakly decreasing in $\ell$; if the game has strategic substitutes ($\beta<0$), then $\operatorname{Var}( \bm{u}^\ell(\bm{G}) \cdot \bm{b}^*)$ is weakly increasing in $\ell$.
		\item[2.] Suppose the planner dislikes variance (i.e., $w<0$). If the game has strategic complements ($\beta>0$), then $\operatorname{Var}( \bm{u}^\ell(\bm{G}) \cdot \bm{b}^*)$ is weakly increasing in $\ell$; if the game has strategic substitutes ($\beta<0$), then $\operatorname{Var}( \bm{u}^\ell(\bm{G}) \cdot \bm{b}^*)$ is weakly decreasing in $\ell$.
	\end{itemize}
\end{proposition}

We now provide the intuition for Proposition \ref{Pr:AV}. Shocks to individuals' standalone marginal returns create variability in the players' equilibrium actions. The assumption that the intervention is mean neutral (part (a) of Assumption \ref{as:K-rotation}) leaves the planner to control only the variances and covariances of these marginal returns with her intervention. Hence, the solution to the intervention problem describes what the planner should do to induce volatilities in actions that maximize the ex-ante expected welfare.

Suppose first that investments are strategic complements. Then a perfectly correlated shock in individual standalone marginal returns is amplified by strategic interaction. In fact, the type of shock that is most amplifying (at a given size) is the one that is perfectly correlated across individuals, with the magnitude of a given individual's shock proportional to the first principal component (his eigenvector centrality). These shocks are exactly what  $\underline{b}^*_{1}=\bm{u}^{1}(\bm{G})\cdot\bm{b}^*$ captures. Hence, this is the dimension of volatility that the planner most wants to increase if she likes variance in actions ($w>0$) and most wants to decrease if she dislikes variance in actions ($w<0$).

If investments are strategic substitutes, then a perfectly correlated shock does not create a lot of variance in actions: The first-order response of all individuals to an increase in their standalone marginal returns is to increase investment, but that in turn makes all individuals decrease their investment somewhat because of the strategic substitutability with their neighbors. Hence, highly positively correlated shocks do not translate into high volatility. The shock profiles that create most variability in actions are the ones in which neighbors have \emph{negatively} correlated shocks. A planner that likes variability in actions will then prioritize such shocks. Because the last eigenvector of the system is correlated with those shocks that have opposite effects on neighbors, this is exactly the type of volatility that is of greatest concern, and this is what the planner will focus on most.

\begin{example}[Illustration in the case of the circle]
Figure \ref{Figure:principalcomponents} depicts six of the eigenvectors/principal components of a circle network with $14$ nodes. The first principal component is a positive vector and so $\mathcal{B}$ projected on $\bm{u}^{1}(\bm{G})$ captures positively correlated shocks across all players. The second principal component (top left panel of Figure \ref{Figure:principalcomponents}) splits the graph into two sides, one with positive entries and the other with negative entries. Hence, $\mathcal{B}$ projected on $\bm{u}^{2}(\bm{G})$ captures shocks that are highly positively correlated on each side of the circle network, with the two opposite sides of the circle being anti-correlated. As we move along the sequence, we can see that $\mathcal{B}$ projected on the $\ell^{\text{th}}$ eigenvector represents shocks that are more and more local. At the extreme,  $\mathcal{B}$ projected on $\bm{u}^{14}(\bm{G})$, (bottom-right panel of Figure \ref{Figure:principalcomponents}) captures the component of shocks that is perfectly anti-correlated across neighbors. 
\end{example}
	


\section{Concluding remarks}\label{sec:concludingremarks}
We study the problem of a planner who seeks to optimally target incentive changes in a network game. Our framework allows for a broad class of strategic and non-strategic spillovers across neighbors. The main contribution of the paper is methodological: we show that
principal components of the network of interaction provide a useful basis for analyzing the effects of an intervention. This decomposition leads to our main result: there is a close relation between the nature of the game (complements or substitutes) and the weight that different principal components receive in the optimal intervention. To develop these ideas in the simplest way, we have focused on a model in which the matrix of interaction is symmetric, the costs of intervention are quadratic, and the intervention itself takes the form of altering the standalone benefits. In the Online Appendix we relax these restrictions and develop extensions of our approach to non-symmetric matrices of interaction, to more general costs of intervention, and to environments where interventions occur via monetary incentives for activity. We also relax Property A, a technical condition which facilitated our basic analysis, and cover a more general class of externalities.


We briefly mention two further applications. In some circumstances, the planner seeks a budget-balanced tax/subsidy scheme in order to improve the economic outcome. In an oligopoly market, for example, a planner could tax some suppliers, thereby increasing their marginal costs, and then use that tax revenue to subsidize other suppliers. The planner will solve a problem similar to the one we have studied here, with the important difference that she will face a different constraint, namely, a budget-balance constraint. In ongoing work, \citet{Galeottietal2018} show that the principal component approach that we employed in this paper is useful in deriving the optimal taxation scheme and, in turn, in determining the welfare gains that can be achieved in supply chains.

We have focused on interventions that alter the standalone marginal returns of individuals. Another interesting problem is the study of interventions that alter the matrix of interaction. We hope this paper stimulates further work along these lines.

\newpage

\bibliographystyle{ecta}
\bibliography{svd}

\appendix

\section{Proofs} \label{sec:proofs}

\begin{proof}[Proof of Theorem \ref{Theorem1}]
We wish to solve
\begin{eqnarray*}
\max_{\bm{b}} &&w \bm{a}^{\tr}\bm{a}\\
\text{s.t.:} && [\bm{I}-\beta\bm{G}]\bm{a}^{*}=\bm{b},\\
&& \sum_i(b_i-\hat{b}_i)^2\leq C.
\end{eqnarray*}
We transform the maximization problem into the basis given by the principal components of $\bm{G}$. To this end, we first rewrite the cost and the objective in the principal components basis, using the fact that norms do not change under the orthogonal transformation $\bm{U}^\tr$. (The norm symbol $\Vert \cdot \Vert$ always refers to the Euclidean norm.) Letting $\bm{y} = \bm{b}-\hat{\bm{b}}$,
\[
K(\bm{b},\hat{\bm{b}})=\sum_i y_i ^2=\Vert \bm{y} \Vert^{2}_{2}=\sum_\ell\underline{y}_\ell^2
\]
and
\[
w \bm{a}^{\tr}\bm{a}=w \Vert\bm{a}\Vert^{2}=w \Vert\underline{\bm{a}}\Vert^{2}=w \underline{\bm{a}}^{\tr}\underline{\bm{a}}.
\]
By recalling that, in equilibrium, $\underline{\bm{a}}^*=[\bm{I}-\beta\bm{\Lambda}]^{-1}\underline{\bm{b}}$, and using the definition $\alpha_\ell=\frac{1}{(1-\beta\lambda_\ell(\bm{G}))^2}$, the intervention problem (IT) can be rewritten as:
\begin{align*}
\max_{\underline{\bm{b}}} \qquad &  w\sum_\ell \alpha_\ell \underline{b}^2_\ell \tag{IT-PC}\\
\text{s.t.} \qquad  &\sum_\ell \underline{y}_\ell^2\leq C.
\end{align*}
We now transform the problem so that the control variable is $\bm{x}$ where $x_{\ell}=y_\ell/\underline{\hat{b}}_\ell$. We obtain
\begin{eqnarray*}
\max_{\bm{x}} &&w\sum_\ell \alpha_\ell (1+x_\ell)^2\underline{\hat{b}}_{\ell}^2 \\
\text{s.t.} &&\sum_\ell \underline{\hat{b}}_\ell^2x_\ell^2\leq C
\end{eqnarray*}

Note that, for all $\ell$, $\alpha_\ell$ are well-defined (by Assumption 1) and strictly positive (by genericity of $\bm{G}$). This has two implications.\footnote{Note that if Assumption 3 does not hold (that is, $w<0$ and $\sum_\ell \underline{\hat{b}}_\ell^2\leq C$) then the optimal solution is $x^*_\ell=-1$ for all $\ell$. This is what we ruled out with Assumption \ref{Ass3}, before Theorem 1.}

First, at the optimal solution $\bm{x}^*$ the resource constraint problem must bind. To see this, note that Assumption 3 says that either $w>0$, or $w<0$ and $\sum_\ell \underline{\hat{b}}_\ell^2>C$. Suppose that at the optimal solution the constraint does not bind. Then, without violating the constraint, we can slightly increase or decrease any $x_\ell$. If $w>0$ (resp. $w<0$) the increase or the decrease is guaranteed to increase (resp. decrease) the corresponding $(x_\ell+1)^2$ (since the $\alpha_\ell$ are all strictly positive).

Second, we show that the optimal solution $\bm{x}^*$ satisfies $x^*_\ell\geq 0$ for every $\ell$ if $w>0$, and $x^*_\ell\in[-1,0]$ for every $\ell$ if $w<0$. Suppose $w>0$ and, for some $\ell$,  $x^*_\ell<0$. Then $[-x^*_\ell+1]^2>[x^*_\ell+1]^2$. Since $w>0$ and every $\alpha_\ell$ is positive, we can raise the aggregate utility without changing the cost by flipping the sign of $x^*_\ell$. Analogously, suppose $w<0$. It is clear that if $x^*_\ell<-1$, then by setting $x_\ell=-1$ the objective improves and the constraint is relaxed; hence, at the optimum, $x^*_\ell\geq-1$. Suppose next that $x_\ell>0$ for some $\ell$. Then $[-x^*_\ell+1]^2<[x^*_\ell+1]^2$. Since $w<0$ and every $\alpha_\ell$ is positive, we can improve the value of the objective function without changing the cost by flipping the sign of $x^*_\ell$.

We now complete the proof. Observe that the Lagrangian corresponding to the maximization problem is
\[
\mathcal{L}=w\sum_\ell \alpha_\ell (1+x_{\ell})^2\underline{\hat{b}}_{\ell}+\mu\left[C-\sum_\ell \underline{\hat{b}}_\ell^2x_\ell^2\right].
\]
Taking our observation above that the constraint is binding at $\bm{x}=\bm{x}^*$, together with the standard results on the Karush--Kuhn--Tucker conditions, the first-order conditions must hold exactly at the optimum with a positive $\mu$:
\begin{equation}\label{Lagrangian}
0=\frac{\partial \mathcal{L}}{\partial x_\ell}=2\underline{\hat{b}}^2_\ell\left[w\alpha_\ell(1+x^*_\ell)-\mu x^*_\ell\right]=0.
\end{equation}
We take a generic $\hat{\bm{b}}$ such that $\underline{\hat{b}}_\ell\neq 0$ for all $\ell$. If for some $\ell$ we had $\mu=w\alpha_\ell$ then the right-hand side of the second equality in (\ref{Lagrangian}) would be $2\underline{\hat{b}}^2_\ell w\alpha_\ell$, which, by the generic assumption we just made and the positivity of $\alpha_\ell$, would contradict (\ref{Lagrangian}). Thus, the following holds with a nonzero denominator:
\[
x^*_\ell=\frac{w \alpha_\ell}{\mu-w \alpha_\ell},
\]
and the Lagrange multiplier $\mu$ is therefore pinned down by
\[
\sum_\ell w^2 \underline{\hat{b}}_\ell^2 \left(\frac{ \alpha_\ell}{\mu-w \alpha_\ell}\right)^2=C.
\]
Note finally that
\[
\rho(\bm{y}^*,\bm{u}^\ell(\bm{G}))=\frac{\bm{y}^*\cdot \bm{u}^\ell(\bm{G})}{\Vert\bm{y}^*\Vert\Vert\bm{u}^\ell(\bm{G})\Vert}=\frac{\underline{y}^*_\ell}{\sqrt{C}}=\frac{\underline{\hat{b}}_\ell x^*_\ell}{\sqrt{C}}=\frac{\Vert\hat{\bm{b}}\Vert}{\sqrt{C}} \rho(\hat{\bm{b}},\bm{u}^\ell(\bm{G}))x^*_\ell\propto_\ell \rho(\hat{\bm{b}},\bm{u}^\ell(\bm{G}))x^*_\ell.
\]
\end{proof}
\begin{proof}[Proof of Proposition \ref{result:smallC}] Part 1. From expression \ref{LM} of Theorem \ref{Theorem1}, it follows that if $C\to 0$ then $\mu \to \infty$. The result follows by noticing that
\[
\frac{r^*_\ell}{r^*_{\ell'}}=\frac{\alpha_\ell}{\alpha_{\ell'}}\frac{\mu-w \alpha_\ell'}{\mu-w \alpha_{\ell}}.
\]

\noindent Part 2. Suppose that $\beta>0$. Using the derivation of the last part of the proof of Theorem 1, we write:
\[
\rho(\bm{y}^*,\bm{u}^\ell(\bm{G}))=\frac{\Vert\hat{\bm{b}}\Vert}{\sqrt{C}} \rho(\hat{\bm{b}},\bm{u}^\ell(\bm{G}))x^*_\ell,
\]
with $x^*_\ell=\frac{w\alpha_\ell}{\mu-w\alpha_\ell}$. From expression \ref{LM} of Theorem \ref{Theorem1}, it follows that if $C\to \infty$ then $\mu \to w\alpha_1$. This implies that $x^*_\ell\to \frac{\alpha_\ell}{\alpha_1-\alpha_\ell}$ for all $\ell\neq 1$. As a result, if $C\to\infty$ then $\rho(\bm{y}^*,\bm{u}^\ell(\bm{G}))\to 0$ for all $\ell\neq 1$. Furthermore, we can rewrite expression \ref{LM} of Theorem \ref{Theorem1} as
\[
\sum_\ell \left(\Vert\hat{\bm{b}}\Vert\rho(\hat{\bm{b}},\bm{u}^\ell(\bm{G})) \frac{x^*_\ell}{\sqrt{C}}\right)^2=1,
\]
and therefore
\[
\lim_{C\to\infty} \sum_\ell \left(\Vert\hat{\bm{b}}\Vert\rho(\hat{\bm{b}},\bm{u}^\ell(\bm{G})) \frac{x^*_\ell}{\sqrt{C}}\right)^2=\lim_{C\to\infty} \left(\Vert\hat{\bm{b}}\Vert\rho(\hat{\bm{b}},\bm{u}^1(\bm{G})) \frac{x^*_1}{\sqrt{C}}\right)^2=1,
\]
where the first equality follows because $x^*_\ell\to \frac{\alpha_\ell}{\alpha_1-\alpha_\ell}$ for all $\ell\neq 1$. The proof for the case of $\beta<0$ follows the same steps, with the only exception that if $C\to\infty$ then $\mu\to w\alpha_n$.
\end{proof}
\begin{proof}[Proof of Proposition \ref{Prop:LargeBudget}] We first prove the result on welfare and then turn to the result on cosine similarity.

\noindent \textbf{Welfare.} Consider the case of strategic complementarities, $\beta>0$. Define by $\tilde{\bm{x}}$ the simple intervention, and note that $\tilde{x}_1=\sqrt{C}/\underline{\hat{b}}_1$ and that $\tilde{x}_{\ell}=0$ for all $\ell>1$. The aggregate utility obtained under the simple intervention is:
\[
W^s=\sum_\ell \underline{\hat{b}}^2_\ell \alpha_\ell (1+\tilde{x}_\ell)^2=\underline{\hat{b}}^2_1 \alpha_1 \tilde{x}_1(\tilde{x}_1+2)+\sum_\ell \alpha_\ell\underline{\hat{b}}^2_\ell.
\]
The aggregate utility at the optimal intervention is
\[
W^*=\sum_\ell \underline{\hat{b}}^2_\ell \alpha_\ell (1+x^*_\ell)^2=\underline{\hat{b}}^2_1 \alpha_1 x^*_{1}(x^*_1+2)+\sum_{\ell\neq 1} \underline{\hat{b}}^2_\ell \alpha_\ell x^*_{\ell}(x^*_\ell+2)+\sum_\ell \alpha_\ell\underline{\hat{b}}^2_\ell
\]
Hence
\begin{align*}
\frac{W^*}{W^s}&= \frac{\underline{\hat{b}}^2_1 \alpha_1 x^*_{1}(x^*_1+2)+\sum_\ell\alpha_\ell \underline{\hat{b}}^2_\ell}{\underline{\hat{b}}^2_1 \alpha_1 \tilde{x}_1(\tilde{x}_1+2)+\sum_\ell \alpha_\ell\underline{\hat{b}}^2_\ell}+\frac{\sum_{\ell\neq 1} \underline{\hat{b}}^2_\ell \alpha_\ell x^*_{\ell}(x^*_\ell+2)}{\underline{\hat{b}}^2_1 \alpha_1 \tilde{x}_1(\tilde{x}_1+2)+\sum_\ell \alpha_\ell \underline{\hat{b}}^2_\ell}\\
&\leq 1+\frac{\sum_{\ell\neq 1} \underline{\hat{b}}^2_\ell \alpha_\ell x^*_{\ell}(x^*_\ell+2)}{\underline{\hat{b}}^2_1 \alpha_1 \tilde{x}_1(\tilde{x}_1+2)+\sum_\ell \alpha_\ell\underline{\hat{b}}^2_\ell}  && \text{\footnotesize as $\tilde{x}_1 \geq x_1^*$} \\
&\leq1+\frac{\sum_{\ell\neq 1} \underline{\hat{b}}^2_\ell \alpha_\ell x^*_{\ell}(x^*_\ell+2)}{\underline{\hat{b}}^2_1 \alpha_1 \tilde{x}^2_1} && \text{\footnotesize summands in denominator are positive} \\&= 1+\frac{\sum_{\ell\neq 1} \underline{\hat{b}}^2_\ell \alpha_\ell x^*_{\ell}(x^*_\ell+2)}{\alpha_1 C} && \text{\footnotesize $\underline{b}_1^2\tilde{x}^2_1=C$; see below}\\ &\leq1+\frac{2\alpha_1-\alpha_2}{\alpha_1}\frac{\Vert\hat{\bm{b}}\Vert^2}{C}\left(\frac{\alpha_2}{\alpha_1-\alpha_2}\right)^2 && \text{\footnotesize see calculation below}\\ &\leq1+\frac{2\Vert\hat{\bm{b}}\Vert^2}{C}\left(\frac{\alpha_2}{\alpha_1-\alpha_2}\right)^2.
\end{align*}
The fact $\underline{b}_1^2\tilde{x}^2_1=C$, used above, follows because the simple policy allocates the entire budget to changing $\underline{b}_1$. The inequality after that statement follows because
\begin{align*}
\sum_{\ell\neq 1} \underline{\hat{b}}^2_\ell \alpha_\ell x^*_{\ell}(x^*_\ell+2)&\leq \alpha_2\sum_{\ell\neq 1} \underline{\hat{b}}^2_\ell x^*_{\ell}(x^*_\ell+2) && \text{ordering of the $\alpha_\ell$} \\ &\leq\alpha_2x^*_{2}(x^*_2+2)\sum_{\ell\neq 1} \underline{\hat{b}}^2_\ell && \text{Corollary 1} \\ &\leq \alpha_2\frac{w \alpha_2}{\mu-w\alpha_2}\left(\frac{w \alpha_2}{\mu-w\alpha_2}+2\right)\sum_{\ell\neq 1} \underline{\hat{b}}^2_\ell && \text{Theorem 1}\\
&\leq\alpha_2\frac{w \alpha_2}{w\alpha_1-w\alpha_2}\left(\frac{w \alpha_2}{w\alpha_1-w\alpha_2}+2\right) \Vert\underline{\hat{\bm{b}}}\Vert^2 \\
&=\left(\frac{\alpha_2}{\alpha_1-\alpha_2}\right)^2\left(2\alpha_1-\alpha_2\right)\Vert\hat{\bm{b}}\Vert^2
\end{align*}
Hence, the inequality
\[
C>\frac{2\Vert\hat{\bm{b}}\Vert^2}{\epsilon}\left(\frac{\alpha_2}{\alpha_1-\alpha_2}\right)^2
\]
is sufficient to establish that $\frac{W^*}{W^s}<1+\epsilon$.  The proof for the case of strategic substitutes follows the same steps; the only difference is that we use $\alpha_n$ instead of $\alpha_1$ and $\alpha_{n-1}$ instead of $\alpha_2$.

\noindent \textbf{Cosine similarity.} We now turn to the cosine similarity result. We focus on the case of strategic complements. The proof for the case of strategic substitutes is analogous. We start by writing a useful explicit expression for
	\begin{equation}
\label{eq:usefulrho}		\rho(\Delta \bm{b}^*,\sqrt{C}\bm{u}^1)=\frac{(\bm{b}^*-\hat{\bm{b}})\cdot (\sqrt{C}\bm{u}^1)}{\Vert\bm{b}^*-\hat{\bm{b}}\Vert \Vert \sqrt{C}\bm{u}^1\Vert}=\frac{(\bm{b}^*-\hat{\bm{b}})\cdot (\bm{u}^1)}{\sqrt{C}},
	\end{equation}
where the last equality follows because, at the optimum, $\Vert\bm{b}^*-\hat{\bm{b}}\Vert^2=C$. At the optimal intervention, by Theorem 1,
	\[ \underline{b}^*_{\ell}-\underline{\hat{b}}_\ell=\frac{w\alpha_\ell}{\mu-w\alpha_\ell}\underline{\hat{b}}_{\ell};
	\]
	now, using the definition $\underline{\bm{b}}=\bm{U}^{T}\bm{b}$, we have that
	\[
	b^*_{i}-\hat{b}_i=w\sum_\ell u_{\ell}^{i}\frac{\alpha_\ell}{\mu-w\alpha_\ell}\underline{\hat{b}}_{\ell}
	\]
	and therefore
	\begin{eqnarray*}
	(\bm{b}^*-\hat{\bm{b}})\cdot \bm{u}^1&=&\sum_i\sum_\ell u_{i}^{1}u_{i}^{\ell}\frac{w\alpha_\ell}{\mu-w\alpha_\ell}\underline{\hat{b}}_{\ell}=\sum_\ell \frac{w\alpha_\ell}{\mu-w\alpha_\ell}\underline{\hat{b}}_{\ell} (\bm{u}^1\cdot \bm{u}^{\ell})=\frac{w\alpha_1}{\mu-w\alpha_1}\underline{\hat{b}}_{1}
	\end{eqnarray*}
	Hence, using this in equation \ref{eq:usefulrho}, we can deduce that
	\begin{equation}
		\rho(\Delta \bm{b}^*,\bm{u}^1)=\frac{1}{\sqrt{C}}\frac{w\alpha_1}{\mu-w\alpha_1}\underline{\hat{b}}_{1}\geq\sqrt{1-\epsilon} \quad \text{ iff } \quad 	\left(\frac{w\alpha_1}{\mu-w \alpha_1}\right)^2\underline{\hat{b}}^2_{1}-C(1-\epsilon)\geq 0. \label{ine} \end{equation}

	The following lemma shows that the inequality after the ``if and only if'' follows from our hypothesis that
\[
C>\frac{2\Vert\hat{\bm{b}}\Vert^2}{\epsilon}\left(\frac{\alpha_2}{\alpha_1-\alpha_2}\right)^2,
\] and thus establishing it completes the proof.
\begin{lemma}\label{LemmaGaleotti}
	Assume $$C>\frac{2\Vert\hat{\bm{b}}\Vert^2}{\epsilon}\left(\frac{\alpha_2}{\alpha_1-\alpha_2}\right)^2.$$ Then
	\begin{equation}
	\left(\frac{w \alpha_1}{\mu-w \alpha_1}\right)^2\underline{\hat{b}}^2_{1}\geq C(1-\epsilon)
	\end{equation}
	\end{lemma}

\begin{proof}[Proof of Lemma \ref{LemmaGaleotti}]
	Note that
	\[
	C>\frac{2\Vert\hat{\bm{b}}\Vert^2}{\epsilon}\left(\frac{\alpha_2}{\alpha_1-\alpha_2}\right)^2 \quad \Longrightarrow \quad \epsilon C>\Vert\hat{\bm{b}}\Vert^2\left(\frac{\alpha_2}{\alpha_1-\alpha_2}\right)^2,
	\]
	and therefore
	\begin{equation}
	C(1-\epsilon)<C-\Vert\hat{\bm{b}}\Vert^2\left(\frac{\alpha_2}{\alpha_1-\alpha_2}\right)^2. \label{eq:C1-e_upperbound}
	\end{equation}
	But then we have the following chain of statements, explained immediately after the display:
	\begin{eqnarray*}
		\left(\frac{w \alpha_1}{\mu-w \alpha_1}\right)^2\underline{\hat{b}}^2_{1}-C(1-\epsilon)&\geq&\left(\frac{w \alpha_1}{\mu-w \alpha_1}\right)^2\underline{\hat{b}}^2_{1}-C+\Vert\hat{\bm{b}}\Vert^2\left(\frac{\alpha_2}{\alpha_1-\alpha_2}\right)^2\\
		&=&	\left(\frac{w\alpha_1}{\mu-w\alpha_1}\right)^2\underline{\hat{b}}^2_{1}-\sum_{\ell}\left(\frac{w\alpha_{\ell}}{\mu-w\alpha_\ell}\right)^2\hat{b}^2_{\ell}+\Vert\hat{\bm{b}}\Vert^2\left(\frac{\alpha_2}{\alpha_1-\alpha_2}\right)^2\\
&=&		\Vert\hat{\bm{b}}\Vert^2\left(\frac{\alpha_2}{\alpha_1-\alpha_2}\right)^2-\sum_{\ell\neq 1}\left(\frac{w\alpha_{\ell}}{\mu-w\alpha_\ell}\right)^2\hat{b}^2_{\ell}\\
		&=&\left(\frac{\alpha_2}{\alpha_1-\alpha_2}\right)^2\sum_{\ell}\underline{\hat{b}}^2_{\ell}-\sum_{\ell\neq 1}\left(\frac{w \alpha_{\ell}}{\mu-w \alpha_\ell}\right)^2\hat{b}^2_{\ell}>0.
	\end{eqnarray*}
The first inequality follows from substituting the upper bound on $C(1-\epsilon)$, statement  (\ref{eq:C1-e_upperbound}) above, which we derived from our initial condition on $C$. The equality after that follows by substituting the condition on the binding budget constraint at the optimum, which we derived in Theorem 1. The next equality follows by isolating the term for the first component in the sum and by noticing that that cancels with the first term. The next equality follows by noticing that $\Vert\hat{\bm{b}}\Vert^2=\Vert{\underline{\hat{\bm{b}}}}\Vert^2$. The final inequality follows because, from the facts that $\mu>w\alpha_1$ and that $\alpha_1>\alpha_2>\cdots>\alpha_n$, we can deduce that for each $\ell > 1$
	\[
	\frac{w \alpha_{\ell}}{\mu-w \alpha_\ell}<\frac{w \alpha_{\ell}}{w \alpha_1-w \alpha_\ell}=\frac{\alpha_{\ell}}{\alpha_1-\alpha_\ell}<\frac{\alpha_{2}}{\alpha_1-\alpha_2} \qedhere
	\] 
\end{proof}
This concludes the proof of Proposition \ref{Prop:LargeBudget}.
\end{proof}

\pagestyle{empty}
\includepdf[pages=-,pagecommand={},width=8.5in]{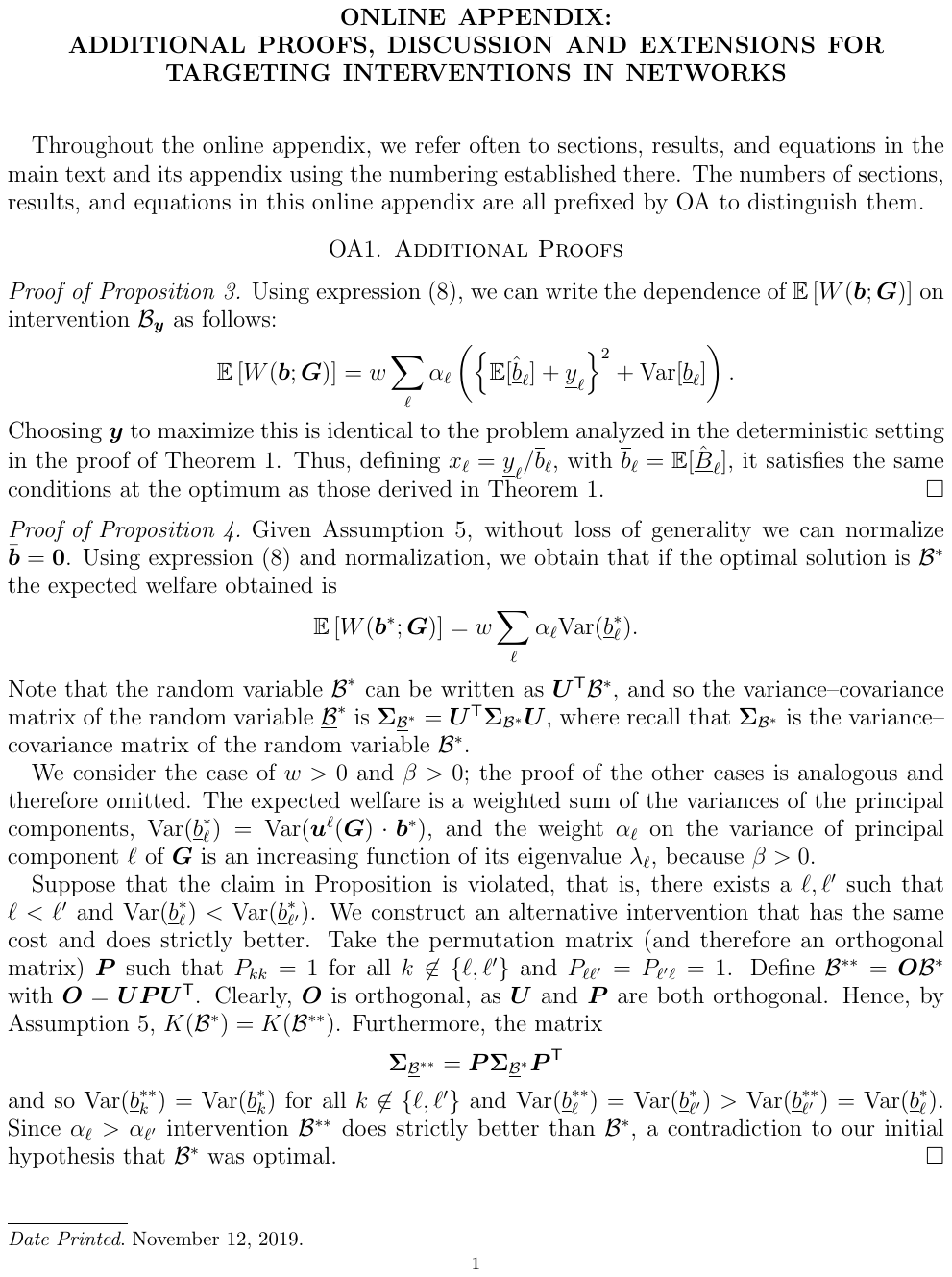}

\end{document}